\documentclass[prb,twocolumn,aps,superscriptaddress,showpacs]{revtex4-2}

\usepackage[english]{babel}

\usepackage[letterpaper,top=2cm,bottom=2cm,left=3cm,right=3cm,marginparwidth=1.75cm]{geometry}

\usepackage{amsmath}
\usepackage{graphicx}
\usepackage{upgreek}
\usepackage{soul}
\usepackage[colorlinks=true, allcolors=blue]{hyperref}
\usepackage{xcolor}

\begin{document}
\title{A frustrated antipolar phase analogous to classical spin liquids
}

\author{G. Bastien}
\affiliation{Charles University, Faculty of Mathematics and Physics, Department of Condensed Matter Physics, Ke Karlovu 5, 121 16 Prague 2, Czech Republic}
\author{D. Rep\v{c}ek}
\affiliation{Institute of Physics, Czech Academy of Sciences, Na Slovance 2, 182 00 Prague, Czech Republic}
\affiliation{Czech Technical University in Prague, Faculty of Nuclear Sciences and Physical Engineering, Department of Solid State Engineering, B\v{r}ehová 7, 115 19 Prague 1, Czech Republic}
\author{A. Eli\'a\v{s}}
\affiliation{Charles University, Faculty of Mathematics and Physics, Department of Condensed Matter Physics, Ke Karlovu 5, 121 16 Prague 2, Czech Republic}
\author{A. Kancko}
\affiliation{Charles University, Faculty of Mathematics and Physics, Department of Condensed Matter Physics, Ke Karlovu 5, 121 16 Prague 2, Czech Republic}
\author{Q. Courtade}
\affiliation{Charles University, Faculty of Mathematics and Physics, Department of Condensed Matter Physics, Ke Karlovu 5, 121 16 Prague 2, Czech Republic}
\author{T. Haidamak}
\affiliation{Charles University, Faculty of Mathematics and Physics, Department of Condensed Matter Physics, Ke Karlovu 5, 121 16 Prague 2, Czech Republic}
\author{M. Savinov}
\affiliation{Institute of Physics, Czech Academy of Sciences, Na Slovance 2, 182 00 Prague, Czech Republic}
\author{V. Bovtun}
\affiliation{Institute of Physics, Czech Academy of Sciences, Na Slovance 2, 182 00 Prague, Czech Republic}
\author{M. Kempa}
\affiliation{Institute of Physics, Czech Academy of Sciences, Na Slovance 2, 182 00 Prague, Czech Republic}
\author{K. Carva}
\affiliation{Charles University, Faculty of Mathematics and Physics, Department of Condensed Matter Physics, Ke Karlovu 5, 121 16 Prague 2, Czech Republic}
\author{M. Vali\v{s}ka}
\affiliation{Charles University, Faculty of Mathematics and Physics, Department of Condensed Matter Physics, Ke Karlovu 5, 121 16 Prague 2, Czech Republic}
\author{P. Dole\v{z}al}
\affiliation{Charles University, Faculty of Mathematics and Physics, Department of Condensed Matter Physics, Ke Karlovu 5, 121 16 Prague 2, Czech Republic}
\author{M. Kratochv\'ilov\'a}
\affiliation{Charles University, Faculty of Mathematics and Physics, Department of Condensed Matter Physics, Ke Karlovu 5, 121 16 Prague 2, Czech Republic}
\author{S. A. Barnett}
\affiliation{ Diamond Light Source, Chilton, Didcot, Oxfordshire, OX11 0DE, United Kingdom}
\author{P. Proschek}
\affiliation{Charles University, Faculty of Mathematics and Physics, Department of Condensed Matter Physics, Ke Karlovu 5, 121 16 Prague 2, Czech Republic}
\author{J. Prokle\v{s}ka}
\affiliation{Charles University, Faculty of Mathematics and Physics, Department of Condensed Matter Physics, Ke Karlovu 5, 121 16 Prague 2, Czech Republic}
\author{C. Kadlec}
\affiliation{Institute of Physics, Czech Academy of Sciences, Na Slovance 2, 182 00 Prague, Czech Republic}
\author{P. Ku\v{z}el}
\affiliation{Institute of Physics, Czech Academy of Sciences, Na Slovance 2, 182 00 Prague, Czech Republic}
\author{R. H. Colman}
\affiliation{Charles University, Faculty of Mathematics and Physics, Department of Condensed Matter Physics, Ke Karlovu 5, 121 16 Prague 2, Czech Republic}
\author{S. Kamba}
\affiliation{Institute of Physics, Czech Academy of Sciences, Na Slovance 2, 182 00 Prague, Czech Republic}

\begin{abstract}

The study of magnetic frustration in classical spin systems was motivated by the prediction and discovery of classical spin liquid states. These uncommon magnetic phases are characterized by a massive degeneracy of their ground state implying a finite magnetic entropy at zero temperature. While the classical spin liquid state was originally predicted in the Ising triangular lattice antiferromagnet in 1950, this state has never been experimentally observed in any triangular magnets. We report here the discovery of an electric analogue of classical spin liquids on a triangular lattice of uniaxial electric dipoles in EuAl$_{12}$O$_{19}$. This new type of frustrated antipolar phase is characterized by a highly degenerate state at low temperature implying an absence of long-range antiferroelectric order, despite short-range antipolar correlations. Its dynamics are governed by a thermally activated process, slowing down upon cooling towards a complete  freezing at zero temperature.

\end{abstract}


\maketitle

\section{Introduction}
\begin{figure*}
\begin{center}
\includegraphics[width=1\linewidth]{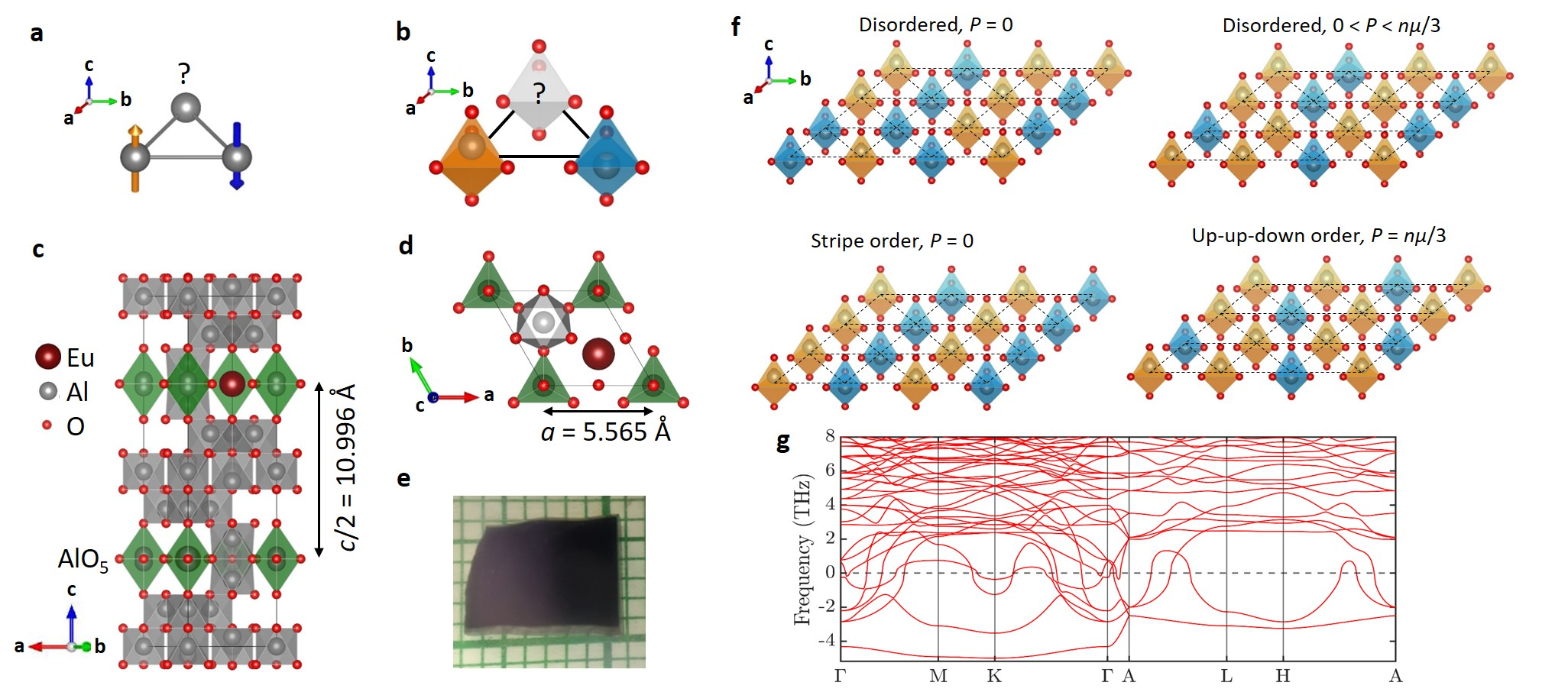}
\caption{\textbf{a} Frustrated magnetism in ITLAFM. It is impossible to place the three spins such that they all point in the opposite direction with respect to their neighbors. \textbf{b} Triangle of Ising-like electric dipoles realizing an equivalent frustration problem. 
The electric dipoles pointing towards $c$ and towards -$c$ are coloured in orange and blue, respectively and the displacement of the Al(5) ion is exaggerated for the clarity of the figure. \textbf{c} View from the $b^\star$ direction of the unit cell of the hexagonal crystal structure of EuAl$_{12}$O$_{19}$ at $T=35\,$K. 
The bipyramid AlO$_{5}$ carrying electric dipoles are highlighted in green. \textbf{d} View from the $c$ axis of a selected part of the unit cell showing the triangular lattice of electric dipoles. \textbf{e} Single crystal of EuAl$_{12}$O$_{19}$ on a millimeter-paper.  \textbf{f} Examples of various electric dipole configurations at the minimum of energy of the ITLAFE. They are characterized by all the triangles harboring one or two dipoles pointing up i.e. there is no triangle with three dipoles pointing in the same direction. Disordered configurations with zero polarization are favored at finite temperature by the minimization of the free energy. The stripe order and the up-up-down order are expected to be selected when the degeneracy is lifted by second nearest neighbor dipolar interactions and by external electric field, respectively~\cite{Takasaki1986, Hwang2008, Wang2014}. \textbf{g} Calculated phonon dispersion curves for EuAl$_{12}$O$_{19}$ with the Al(5) ion sitting at the centre of the AlO$_5$ bipyramid. Only phonons below 8\,THz are shown to focus on the low energy branches. Imaginary frequencies are represented by negative values.}
    \label{Fig1} 
\end{center}
\end{figure*}

Frustrated magnets are systems where all the magnetic interactions cannot be simultaneously satisfied~\cite{Ramirez1994, Balents2010}. In classical spin systems, this frustration can lead to highly degenerate ground states, referred to as classical spin liquid states. These uncommon magnetic phases have been experimentally observed in pyrochlore magnets~\cite{Harris1997, Ramirez1999} and in Kagome magnets~\cite{Kermarrec2021}. The canonical historic example of a classical spin liquid is the type proposed by Wannier in 1950, in the Ising triangular lattice antiferromagnet (ITLAFM)~\cite{Wannier1950}. This model consists of Ising spins arranged in a planar triangular lattice with antiferromagnetic interactions favoring antiparallel alignment of neighboring spins (\textbf{Figure~\ref{Fig1}a}). There is not a single configurational state with lowest magnetic energy, instead many possible energy-degenerate states result in the same lowest energy. This implies an absence of long-range ordering and a finite magnetic entropy at zero temperature in contradiction to the third law of thermodynamics~\cite{Balents2010, Ramirez1999, Wannier1950}. The disordered ground state is topological, since spin configurations can be described in terms of topologically protected strings, which can be formed or broken only upon nucleation of spinons~\cite{Jiang2005, Zhou2022}. However, this classical spin liquid state has never been observed in any ITLAFM. Indeed, experimental realizations of an $S=1/2$ ITLAFM deviate from the strict Ising limit and are better described by the quantum Ising model, where transverse quantum spin fluctuations or an effective transverse field plays a major role~\cite{Arh2022, Gao2022, Chen2024, Moessner2000, Li2020}.

 Several models initially developed to describe magnetism have already motivated the discovery and study of new phenomenon in dielectrics such as electric dipolar glasses~\cite{Courtens1984, Vugmeister1990, Hoechli1990} and ferroelectric quantum criticality~\cite{Rowley2014, Rowley2016}.
The discovery of geometrical frustration in ice crystal~\cite{Pauling1935, Artemov2021} inspired the research on frustrated magnetism in spin ices~\cite{Balents2010, Harris1997, Ramirez1999} and the research on ices of electric dipoles~\cite{Seshadri2006, McQueen2008,Coates2021,  HickoxYoung2022}. The study of quantum magnetism in organic triangular magnets revealed also the presence of a frustrated lattice of charge-displacive electric dipole, which may realize a quantum electric dipole liquid besides the putative quantum spin liquid state.~\cite{Hotta2010, Hassan2018}

 In this paper we propose to explore frustrated magnetism of ITLAFM based on the study of an analogue of the ITLAFM on a lattice of electric dipoles.  Indeed, ion-displacive electric dipoles have the advantage that they can be intrinsically Ising~\cite{Rowley2016, Rensen1969, Shen2014}. The magnetoplumbite crystal structure in particular harbors a triangular lattice of Ising-like electric dipoles formed by an MO$_5$ bipyramid ( M = Al, Ga, Fe)~\cite{Rensen1969, Shen2014, Kimura1990, Iyi1990, Albanese1992, Cao2015, Holtstam2020} (\textbf{Figure~\ref{Fig1}b-d}). The combined effects of a triangular geometry, and interactions favoring antipolar arrangement of neighboring dipoles results in an electric frustration (Figure~\ref{Fig1}b)~\cite{Wang2014, Zhang2020}. Thus the magnetoplumbite crystal structure realizes an electric analog of the $S=1/2$ ITLAFM that we propose to call the Ising triangular lattice antiferroelectric (ITLAFE).

The geometrical frustration of the magnetoplumbite crystal structure  was first noted by  Wang \textit{et al.}~\cite{Wang2014}. Using DFT calculations, they showed that the interaction between electric dipoles is dominated by electric dipole-dipole interactions. As a consequence they favor antipolar alignment of dipoles located on the same triangular layer and polar alignment of electric dipoles on adjacent layers. However experimental and theoretical studies of barium-hexaferrite BaFe$_{12}$O$_{19}$ revealed the important role of quantum fluctuations favoring a quantum paraelectric ground state~\cite{Rowley2016, Shen2014, Cao2015, Zhang2020, Shen2016, Zhang2022}. One way to reduce the influence of quantum fluctuations on the dielectric properties is to change the ion in the double pyramid from Fe$^{3+}$ to the smaller Al$^{3+}$, as theoretically proposed by Wang \textit{et al.}~\cite{Wang2014}. The presence of AlO$_5$ electric dipoles in the hexaaluminates AAl$_{12}$O$_{19}$ (A = Ca, Sr, Pb) has previously been confirmed using x-ray and neutron diffraction~\cite{Kimura1990, Iyi1990, Li2016}, however, low temperature dielectric studies of these materials have been lacking.

\begin{figure*}[t]
\begin{minipage}{0.32\linewidth}
\begin{center}
\includegraphics[width=0.95\linewidth]{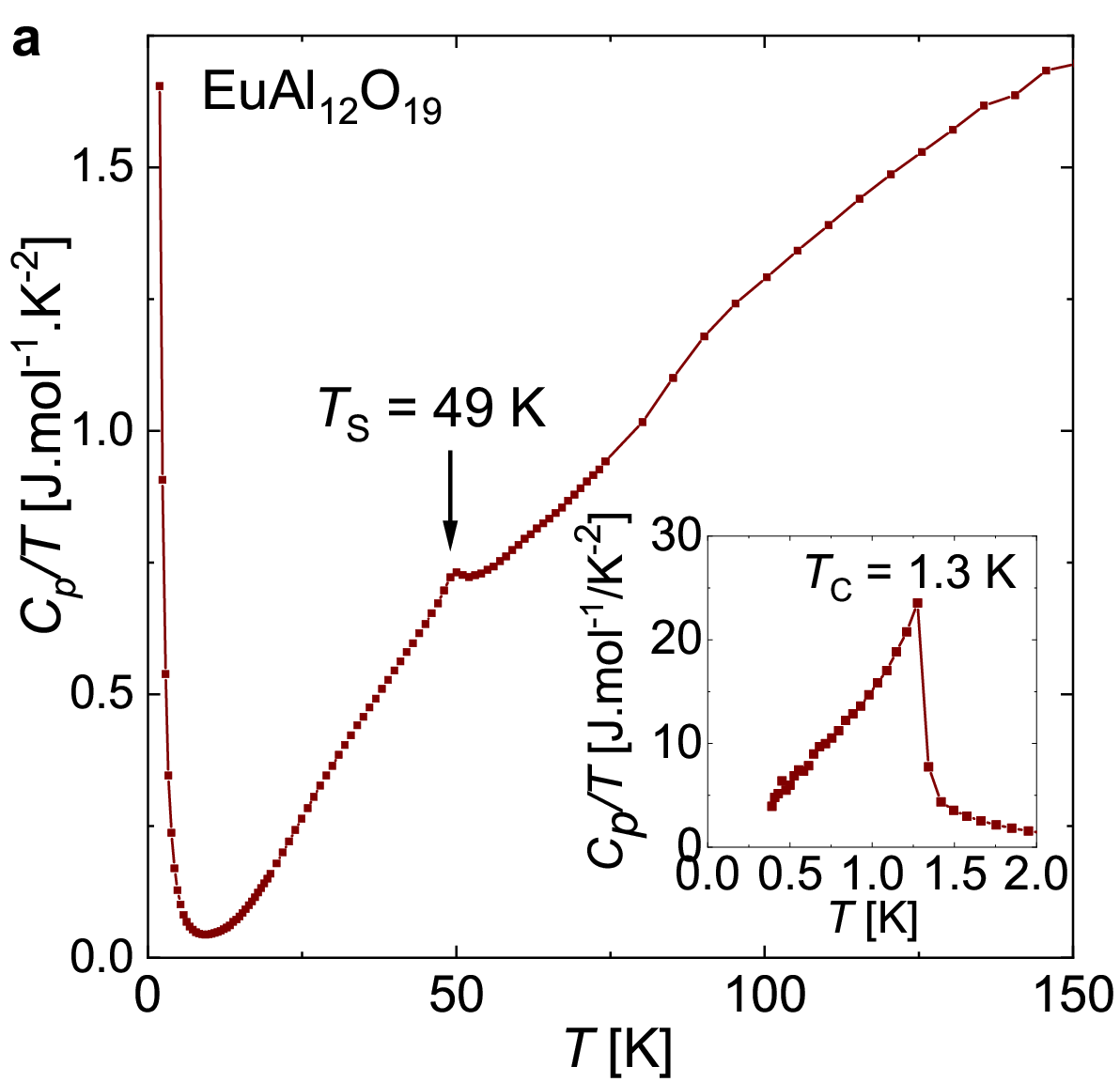}
\end{center}
\end{minipage}
\begin{minipage}{0.32\linewidth}
\begin{center}
\includegraphics[width=0.95\linewidth]{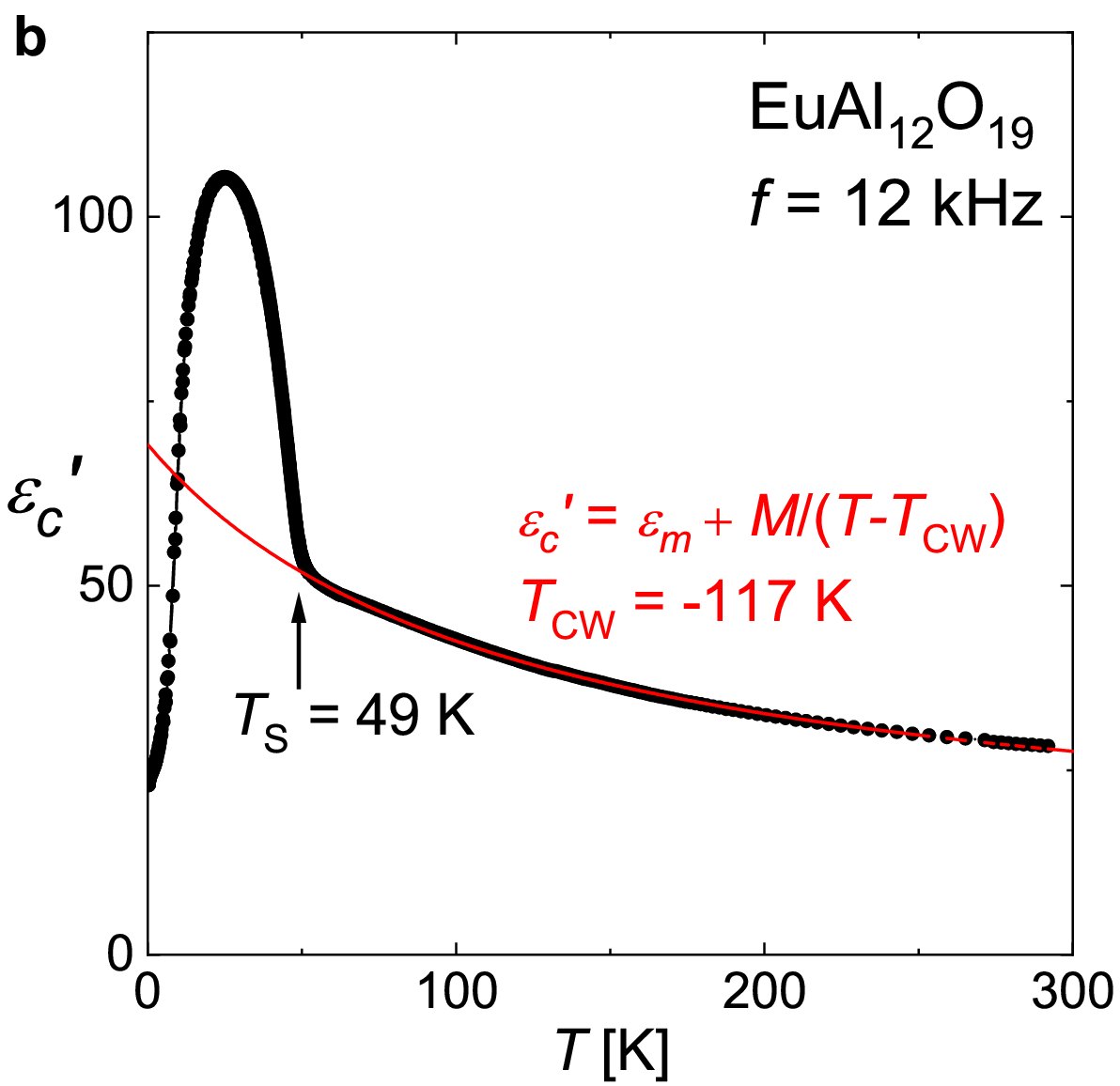}
\end{center}
\end{minipage}
\begin{minipage}{0.32\linewidth}
\begin{center}
\includegraphics[width=0.95\linewidth]{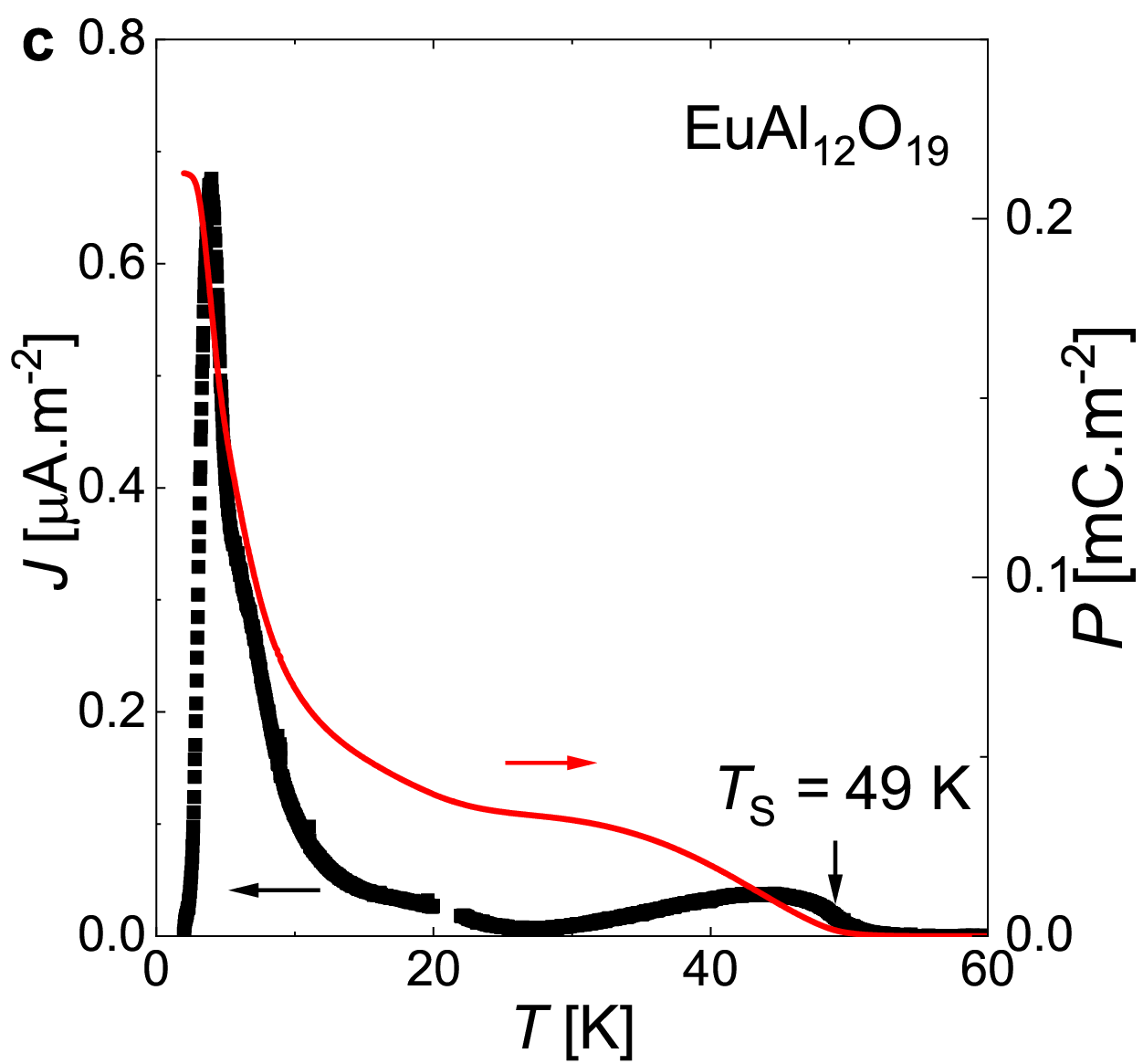}
\end{center}
\end{minipage}
\caption{(a) Specific heat divided by temperature $C_p/T$ of EuAl$_{12}$O$_{19}$ as a function of temperature revealing a second order phase transition at $T_{S}=49$\,K and the ferromagnetic transition at $T_C=1.3$\,K. (b) Dielectric permittivity of EuAl$_{12}$O$_{19}$ along the $c$ axis measured at a frequency of $f=12$\,kHz as a function of temperature. The transition at $T_\mathrm{S} = $49\,K is clearly indicated by an abrupt change of slope. The red curve is a Curie-Weiss fit giving the large negative Curie-Weiss temperature of $T_\mathrm{CW} = -117$\,K. (c) Pyroelectric current in black and polarization in red as a function of temperature. The crystal was initially cooled down to $T=2\,$K under an electric field of $E_\mathrm{DC}=3.2$\,kV/cm along the $c$ axis. Then the pyroelectric current was measured upon heating at $q=1$\,K/min in absence of electric field. It shows a strong reduction of the polarization around $T \approx 4$\,K followed by its suppression around $T_{S}=49$\,K.}
\label{P}
\end{figure*}

In this paper, we reveal that the lattice of electric dipoles in the hexaaluminate EuAl$_{12}$O$_{19}$ is an electric analogue of the classical frustrated spin systems.
EuAl$_{12}$O$_{19}$ is a luminescent material~\cite{Verstegen1974, Stevels1976} and a quasi-two-dimensional ferromagnet below $T_\mathrm{C}=1.3$\,K
~\cite{Bastien2024}. We report the occurrence of a second order phase transition in EuAl$_{12}$O$_{19}$ at $T_\mathrm{S}=49\,$K with a strong enhancement of the dielectric permittivity below this transition. Combining polarization measurements and single crystal synchroton x-ray diffraction we show that this transition does not correspond to any ferroelectric or antiferroelectric ordering. The AlO$_{5}$ electric dipoles build up short range antipolar correlations upon cooling but they remain dynamically disordered at all temperatures due to the geometrical frustration. We observe a soft dielectric relaxation coming from the electric dipoles and we have followed this relaxation from the THz range at room temperature down to the Hz range near $T \approx 5$\,K. The dynamic switching between degenerate dipole configurations is a thermally activated process well described by the Arrhenius law.

Based on our data we propose that the electric dipoles, created by the displacement of Al$^{3+}$ ions within their AlO$_5$ bipyramids, form a frustrated antipolar state. The electric dipoles build up short range correlations upon cooling but they do not form any long-range order, thus the frustrated antipolar state is analogous to ordinary liquids, which lack crystalline order and to classical spin liquids, which lack magnetic order despite strong magnetic correlations~\cite{ Ramirez1999, Harris1997, Balents2010, Kermarrec2021}. Therefore, we propose to name it a classical electric dipole liquid. A few examples of ordered and disordered electric dipole configurations occupying the minimum of energy are given in \textbf{Figure~\ref{Fig1}f}. Disordered dipole configurations are favored by a large  configurational entropy of $S=0.323R$~\cite{Wannier1950}. The system evolves continuously among the manifold of degenerate states, with its dynamics slowing down upon cooling towards a complete freezing only at $T=0\,$K, as predicted for classical spin liquids~\cite{Balents2010, Harris1997, Moessner1998}. This classical electric dipole liquid state differs by its dynamical properties from electric dipolar glasses and spin glasses which occur in disordered systems and freeze at a finite temperature~\cite{Courtens1984, Vugmeister1990, Hoechli1990}. It also differs from quantum electric dipole liquids and quantum spin liquid states which remain dynamic even at zero temperature due to quantum fluctuations~\cite{Balents2010, Hotta2010, Hassan2018,  Shen2016}.


\section{Results and discussion}

\subsection{Antipolar interactions indicated by crystal structure relaxation and
phonon spectrum computation}

We have examined theoretically the presence of interacting electric
dipoles in EuAl$_{12}$O$_{19}$ by computing lattice vibration spectra using DFT+U method (\textbf{Figure~\ref{Fig1}g}). The calculation performed with the Al(5) ion sitting at the center of the AlO$_5$ bipyramid shows a large number of soft modes with imaginary frequencies throughout the Brillouin zone, plotted as negative frequencies in Figure~\ref{Fig1}g. It confirms the instability of the paraelectric crystal structure of EuAl$_{12}$O$_{19}$ described by the unit cell shown in Figure~\ref{Fig1}c (space group $P6_3/mmc$). These results show similarity with phonon calculation in isostructual BaFe$_{12}$O$_{19}$ indicating the same instability~\cite{Wang2014, Zhang2020}. A second calculation was performed with the Al(5) ion in an off-centered position implying a ferroelectric crystal structure (space group $P6_3mc$) (See Figure~S1 and Table~S1 in Supplemental). The two dipoles within the unit cell prefer parallel alignment, implying ferroelectric order along the c axis. Imaginary phonon frequencies were still observed approaching the boundary of the Brillouin
zone in the planar direction indicating that phonon modes with Al(5)
displaced oppositely in the neighboring cell would be
of lowest energy. In order to verify the character of interactions between neighboring polar cells we have also evaluated energies of a cell composed of two unit cells (along the $a$ axis) with the two possible dipole alignments. We found that the antiparallel ordered system has energy lower than the parallel one by 26\,meV per f.u. This interaction clearly prefers antipolar arrangement, and the same is expected along the $b$ axis due to symmetry. Combined with the triangular geometry, these antipolar interactions imply the geometric frustration of the lattice of electric dipoles.

\begin{figure*}[t]
\begin{minipage}{0.49\linewidth}
\begin{center}
\includegraphics[width=0.85\linewidth]{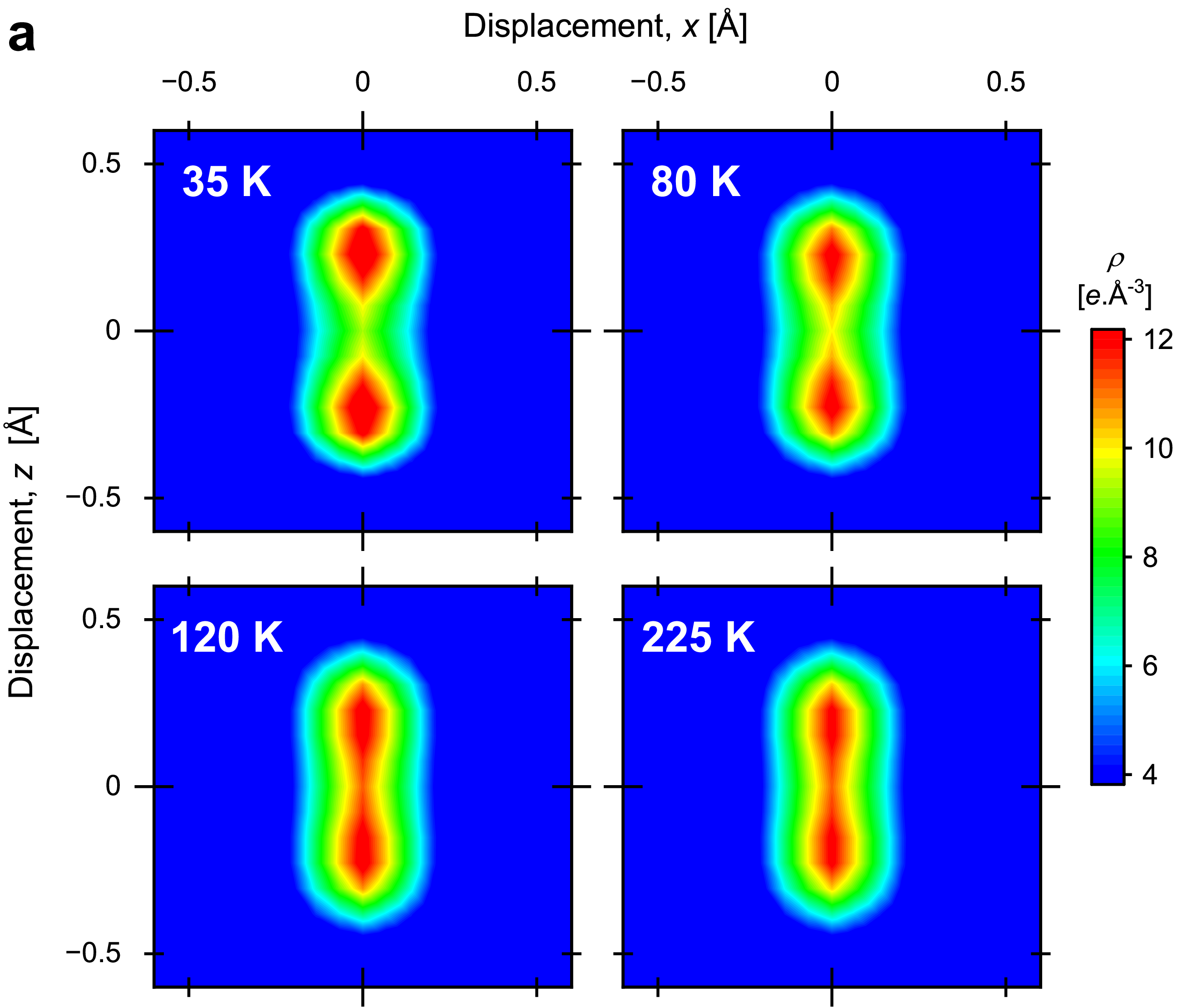}
\end{center}
\end{minipage}
\hfill
\begin{minipage}{0.49\linewidth}
\begin{center}
\includegraphics[width=0.85\linewidth]{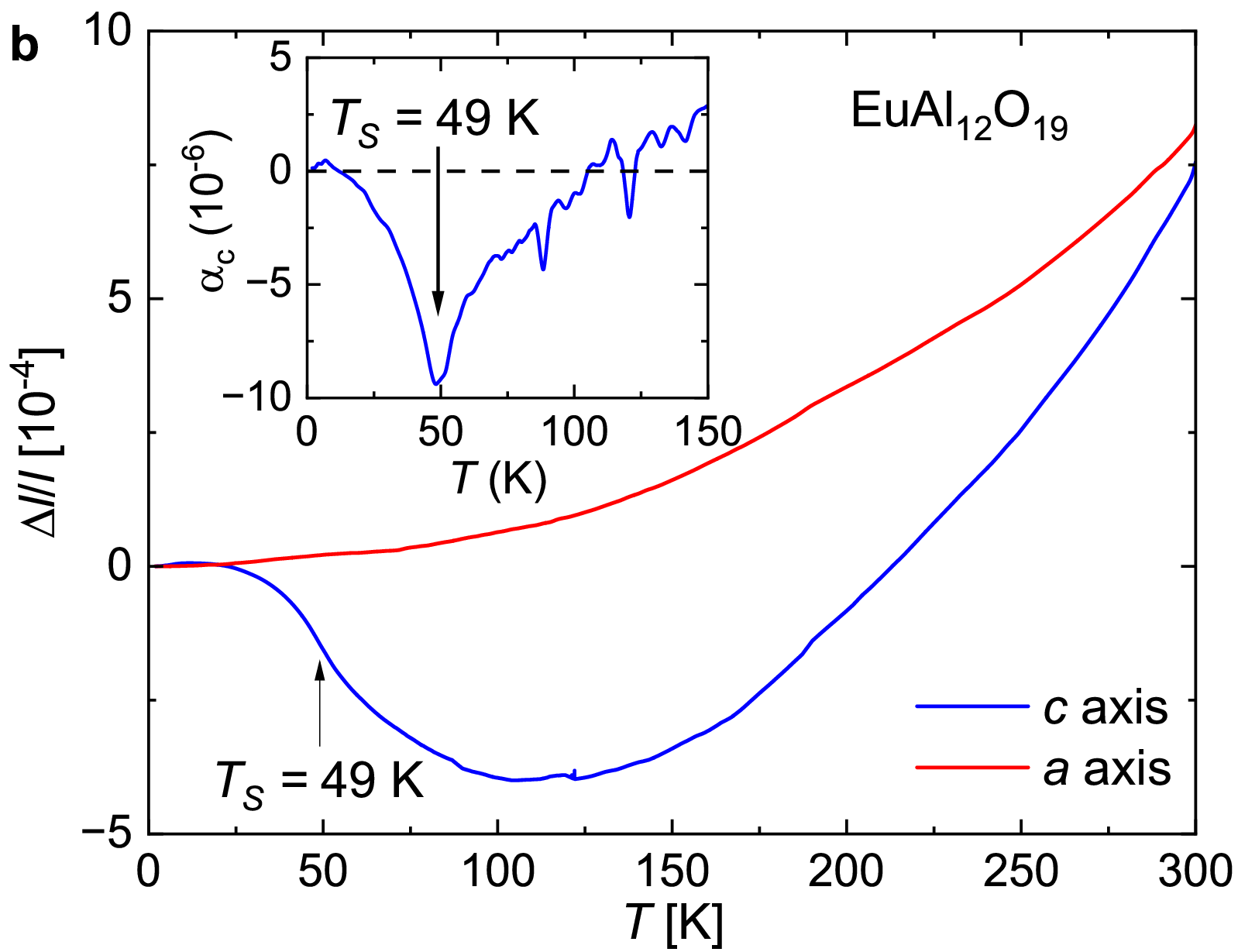}
\end{center}
\end{minipage}
\caption{\textbf{a} Temperature dependence of the electron density distribution obtained from MEM analysis confirming that the Al(5) ion is trapped in a double well potential. The figure consists in 2D-cuts within the (101) plane, centred at (0,0,1/4). \textbf{b} Relative length change of EuAl$_{12}$O$_{19}$ along the $a$ and $c$ axis as a function of temperature. The inset shows the thermal expansion coefficient $\alpha_c$ as a function of temperature.}
\label{Te}
\end{figure*}

\subsection{Unusual phase transition at $T_S=49\,$K}

Specific heat measurements in EuAl$_{12}$O$_{19}$ show a succession of two second order phase transitions at $T_\mathrm{S}=49$\,K and $T_\mathrm{C} = 1.3$\,K as represented in \textbf{Figure~\ref{P}a}. The latter  one  at $T_\mathrm{C} = 1.3$\,K was previously identified as a ferromagnetic ordering of magnetic moments on the Eu$^{2+}$ site~\cite{Bastien2024}. The former transition at $T_\mathrm{S} = 49$\,K is accompanied by an abrupt change of the slope in the temperature dependence of the dielectric permittivity $\varepsilon'$ and the suppression of pyroelectric currents, as shown by \textbf{Figure~\ref{P}b,c}. Such dielectric anomalies are usually associated with improper or pseudoproper ferroelectric transition~\cite{Dvorak1974, Levanyuk1974, Kamba2017}.

At temperature higher than $T_\mathrm{S} = 49$\,K, the dielectric permittivity $\varepsilon'$ follows the Curie-Weiss (C-W) law: $\varepsilon'=\varepsilon_m + M/(T-T_\mathrm{CW})$. The negative Curie-Weiss temperature $T_\mathrm{CW}$ = - 117\,K indicates incipient ferroelectricity~\cite{Rowley2014, Shen2014}. $\varepsilon_m=11.4$ stands for the temperature independent contribution to $\varepsilon'$. The parameter  $M$ = 6750\,K gives us an estimate of the magnitude of the electric dipoles $\mu=0.97\,e$.\AA\, using the equation $M=n\mu^2/k_\mathrm{B}$, where $n$ is the density of electric dipoles and $k_\mathrm{B}$ is the Boltzmann constant~\cite{Shen2014}. Below $T=15$\,K, $\varepsilon'$ drops much below the value given by the C-W law indicating the shift of dielectric relaxation frequency below the measurement frequency.
This result contrasts with the previously studied isostructural compounds SrFe$_{12}$O$_{19}$ and BaFe$_{12}$O$_{19}$, where strong quantum fluctuations prevent any slowing down of the dynamics of the electric dipoles at low temperature~\cite{Shen2014, Cao2015, Rowley2016, Zhang2020}. 

To test whether the transition at $T_\mathrm{S}$ = 49\,K in EuAl$_{12}$O$_{19}$ is a ferroelectric transition, we probed the electrical polarization both with second harmonic generation (SHG) experiments and pyroelectric current methods. SHG was measured both along and perpendicular to the crystallographic $c$ axis in the temperature range from 10\,K to room temperature. No SHG signal was observed neither above or below $T_{\mathrm{S}}=49\,$K confirming that the structure remains centrosymmetric. Measurements of pyroelectric current indicate a weak polarization in poled EuAl$_{12}$O$_{19}$ below $T_{\mathrm{S}}=49\,$K (Figure~\ref{P}c). However it grows linearly with the electric field applied upon cooling without any hint of saturation up to $E_\mathrm{DC}=6\,$kV/cm and it can be interpreted as an electric-field induced polarization $P=\varepsilon_0 \varepsilon'_c E$ (see Figure~S2 in Supplemental). Thus SHG experiment and the pyroelectric current experiments both show that the centre of symmetry of EuAl$_{12}$O$_{19}$ is preserved at all temperatures in the absence of an external electric field and the transition at $T_\mathrm{S} = 49$\,K is not a ferroelectric ordering transition. The observation of a polarization measured after cooling under electric field can be explained by the preferential selection of dipole configurations from the ground-state manifold that have finite polarization (see Figure~\ref{Fig1}f). The temperature dependence of the electrical polarization is rather unusual since we observe its strong reduction above $T=\,$4\,K (Figure~\ref{P}c), which is not directly related to any phase transition. This feature can be interpreted as a dynamical release of the polarization and it will be discussed in more detail later in the context of the relaxation processes detected by THz, microwave and dielectric spectroscopies. 

The strong increase of $\varepsilon'$ below $T_{\mathrm{S}}=49\,$K could in principle come from the contribution of ferroelectric domain walls like in the improper ferroelectric 2H-BaMnO$_3$~\cite{Kamba2017}. We excluded this scenario by the measurement of an independence of $\varepsilon'$ on applied static (DC) electric field, at least up to $E_{\mathrm{DC}}=4.5$\,kV/cm (see Figure~S3 in Supplemental)~\cite{Kumar2013,Kamba2017, Goian2023}. This result further confirms the absence of ferroelectric order in EuAl$_{12}$O$_{19}$.


\subsection{Absence of symmetry change at $T_\mathrm{S}=49$\,K from x-ray diffraction}

To precisely map out structural changes induced upon cooling in  EuAl$_{12}$O$_{19}$, we performed  single crystal x-ray diffraction at temperature down to $T=35$\,K. On cooling from 225 K to 35 K no signatures of any symmetry change, such as the appearance of superstructure peaks, is evident (see Figure~S4 and Table~S II-S V in Supplemental). The best space group to describe the diffraction results remains $P6_3/mmc$ at all temperatures. All reflection extinction conditions of the $P6_3/mmc$ space group are strictly maintained within the resolution of our data down to the lowest measured temperature and attempts to fit with lower symmetry space group such as the ferroelectric space group $P6_3mc$ or the centrosymmetric space group $P3\bar{m}1$ did not lead to any significant improvement of the refinement. These measurements indicate that the electric dipole lattice is disordered at all temperatures. As a consequence the transition at $T_\mathrm{S}=49$\,K does not correspond to the formation of the antiferroelectric stripe order previously predicted by Monte Carlo calculations~\cite{Wang2014} and the nature of this transition remains unclear. 
The distance between the equilibrium position of Al(5) and the center of the bipyramid at $T=225\,$K was estimated at $\delta=0.21$\,\AA\, from structural refinement. The room temperature off-centering displacement extracted from x-ray or neutron diffraction in materials with the magnetoplumbite increases slightly with the ionic radius of the A$^{2+}$ ion from $\delta=0.17$\,\AA\, in CaAl$_{12}$O$_{19}$ to $\delta=0.21$\,\AA\,~\cite{Kimura1990, Iyi1990, Li2016} in EuAl$_{12}$O$_{19}$ and it decreases with the ionic radius of the B$^{3+}$ ion from $\delta=0.21$\,\AA\, in EuAl$_{12}$O$_{19}$ and SrAl$_{12}$O$_{19}$ to $\delta=0.10$\,\AA\, in SrFe$_{12}$O$_{19}$~\cite{Kimura1990, Graetsch1994, Li2016}. While no significant evolution of the off-centering with temperature were measured in hexaaluminates EuAl$_{12}$O$_{19}$ and CaAl$_{12}$O$_{19}$~\cite{Li2016}, the off-centering of the Fe$^{3+}$ ion in BaFe$_{12}$O$_{19}$ gets significantly reduced upon cooling from $\delta =0.177$\,\AA\, at $T=295\,$K down to $\delta =0.098$\,\AA\, at $T=4\,$K~\cite{Albanese1992, Cao2015}. The increased separation distance at low temperature results in a larger barrier to quantum tunneling of the ion and it explains why EuAl$_{12}$O$_{19}$ is much less sensitive to quantum fluctuations than the previously studied BaFe$_{12}$O$_{19}$~\cite{Zhang2020, Shen2016}.


\begin{figure*}[t]
\begin{minipage}{0.49\linewidth}
\begin{center}
\includegraphics[width=0.85\linewidth]{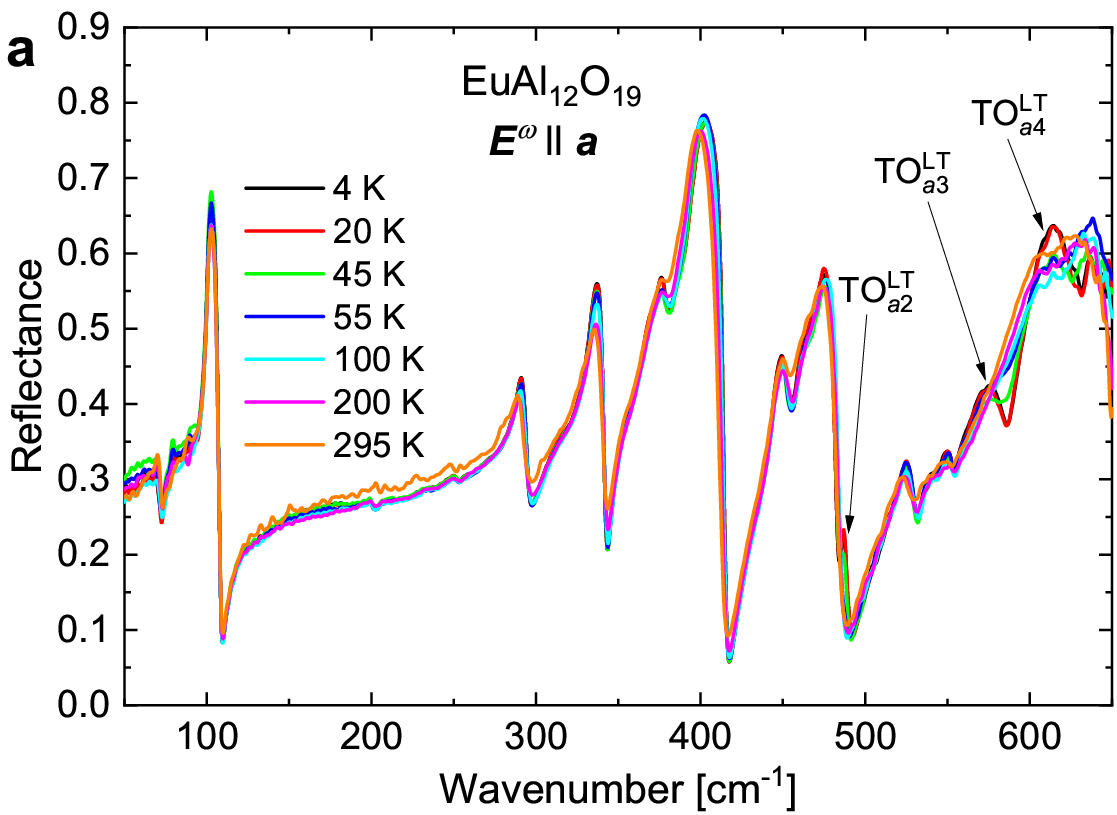}
\includegraphics[width=0.9\linewidth]{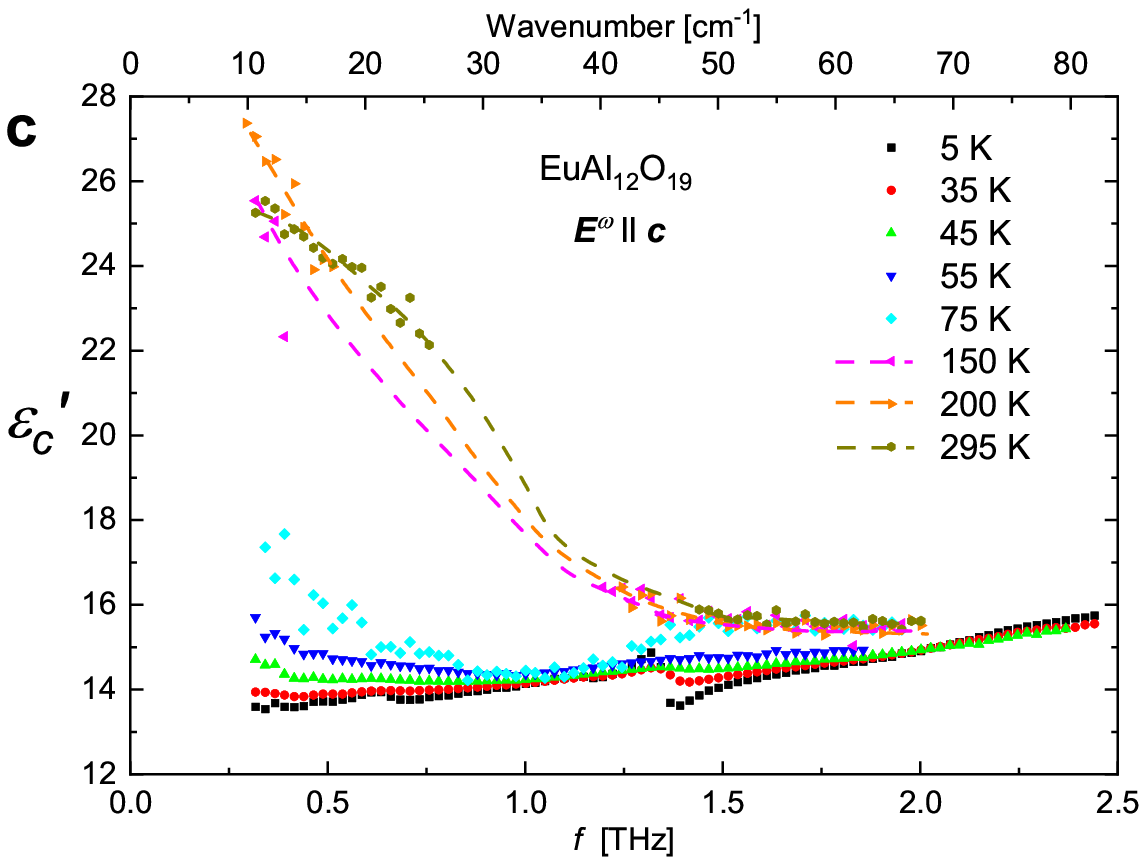}
\end{center}
\end{minipage}
\hfill
\begin{minipage}{0.49\linewidth}
\begin{center}
\includegraphics[width=0.85\linewidth]{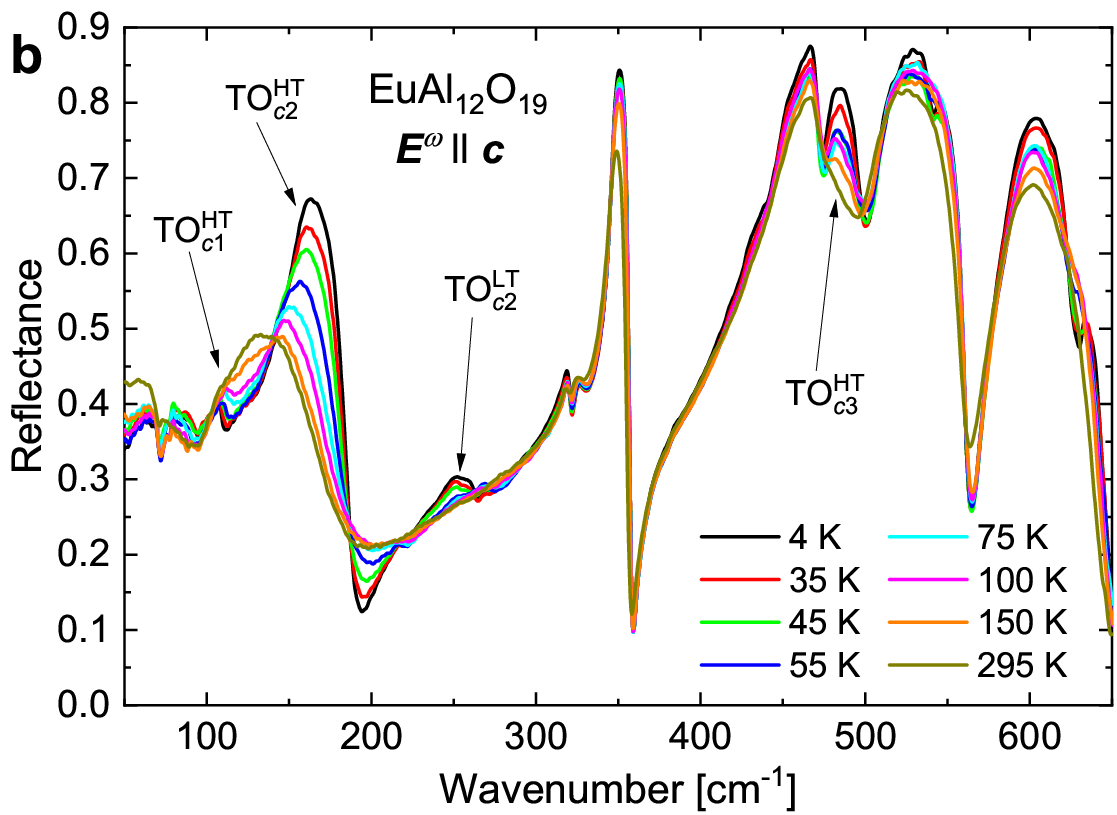}
\includegraphics[width=0.9\linewidth]{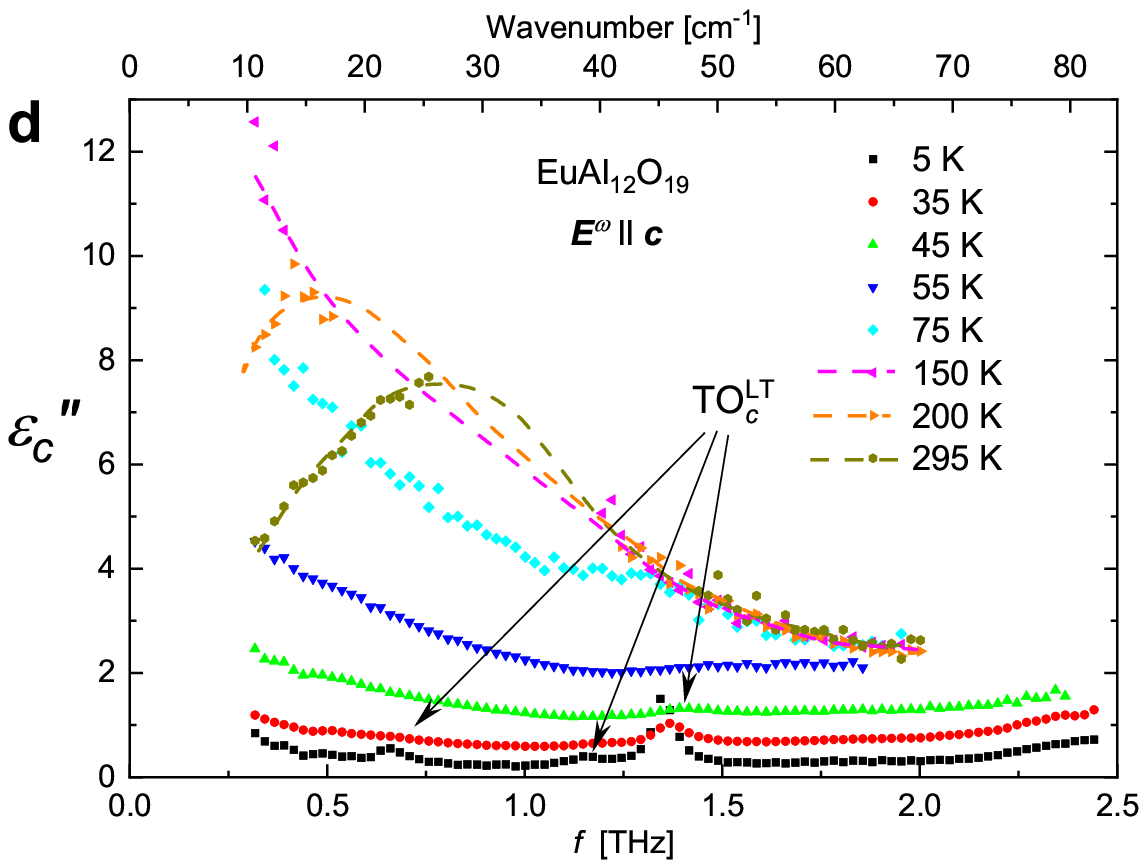}
\end{center}
\end{minipage}
\caption{\textbf{a-b} Temperature evolution of the IR reflectivity spectra of EuAl$_{12}$O$_{19}$ measured in the $\textbf{E}^{\omega}$$\parallel$~$\textbf{a}$ configuration in \textbf{a} and in the $\textbf{E}^{\omega}$$\parallel$~$\textbf{c}$ configuration in \textbf{b}. New transverse optic modes get visible upon cooling, they are pointed by arrows and labeled TO$_{a}^{\mathrm{LT}}$ for $\textbf{E}^{\omega}$$\parallel$~$\textbf{a}$ and TO$_{c}^{\mathrm{LT}}$  for $\textbf{E}^{\omega}$$\parallel$~$\textbf{c}$. Some other modes already visible at room temperature get strongly enhanced upon cooling, they are labelled TO$_{c}^{\mathrm{HT}}$. \textbf{c-d} Frequency dependence of the THz real \textbf{c} and imaginary parts \textbf{d} of the complex permittivity spectra of EuAl$_{12}$O$_{19}$, measured in the $\textbf{E}^{\omega}$$\parallel$~$\textbf{c}$ configuration. A broad relaxation R1 spreads practically across
the whole THz range at $T=295$\,K and its absorption is so
strong that there is even missing gap in the original
transmittance data, filled by dashed line as guide to the eye. TO$_{c}^{\mathrm{LT}}$ indicates three phonon modes appearing below $T_\mathrm{S}=49$\,K.}
\label{IR}
\end{figure*}

To more clearly follow the temperature dependence of the positional distribution of the Al(5) ion within its double pyramid, we performed maximum-entropy method (MEM) analysis on the phased single crystal synchrotron structure factor data. (\textbf{Figure~\ref{Te}a}, Figure~S5 and Figure~S6 in Supplemental). The electron density distribution confirms a double well potential of the Al(5) site, with a distance between the equilibrium position and the center of the bipyramid at $T=35\,$K of $\delta=0.24\,$\AA\, in agreement with $\delta=0.22$\,\AA\, from the structural refinement.

We also observed negative thermal expansion along the $c$ axis within a broad range of temperature from 12\,K to 110\,K both by thermal expansion measurements (\textbf{Figure~\ref{Te}b}) and by single crystal x-ray diffraction (Figure~S7 in Supplemental). The thermal expansion coefficient $\alpha_c$ undergoes a minimum around the transition at $T_\mathrm{S}=49$\,K. This negative thermal expansion
must come from the slowing down of the vibration of the Al(5) ion in the double pyramid~\cite{Mittal2018} and from the building up of antipolar correlations between electric dipoles~\cite{Ramirez2000, Zhao2024}. Thus the observation of negative thermal expansion across a broad temperature range indicates that the formation of antipolar correlations between electric dipoles upon cooling is a continuous process, like the formation of magnetic correlations in frustrated magnets~\cite{Ramirez1994, Ramirez2000, Wang2014}.


\subsection{Dynamical properties of the lattice of electric dipoles}


\begin{figure*}[t]
\begin{minipage}{0.52\linewidth}
\begin{center}
\includegraphics[width=0.9\linewidth]{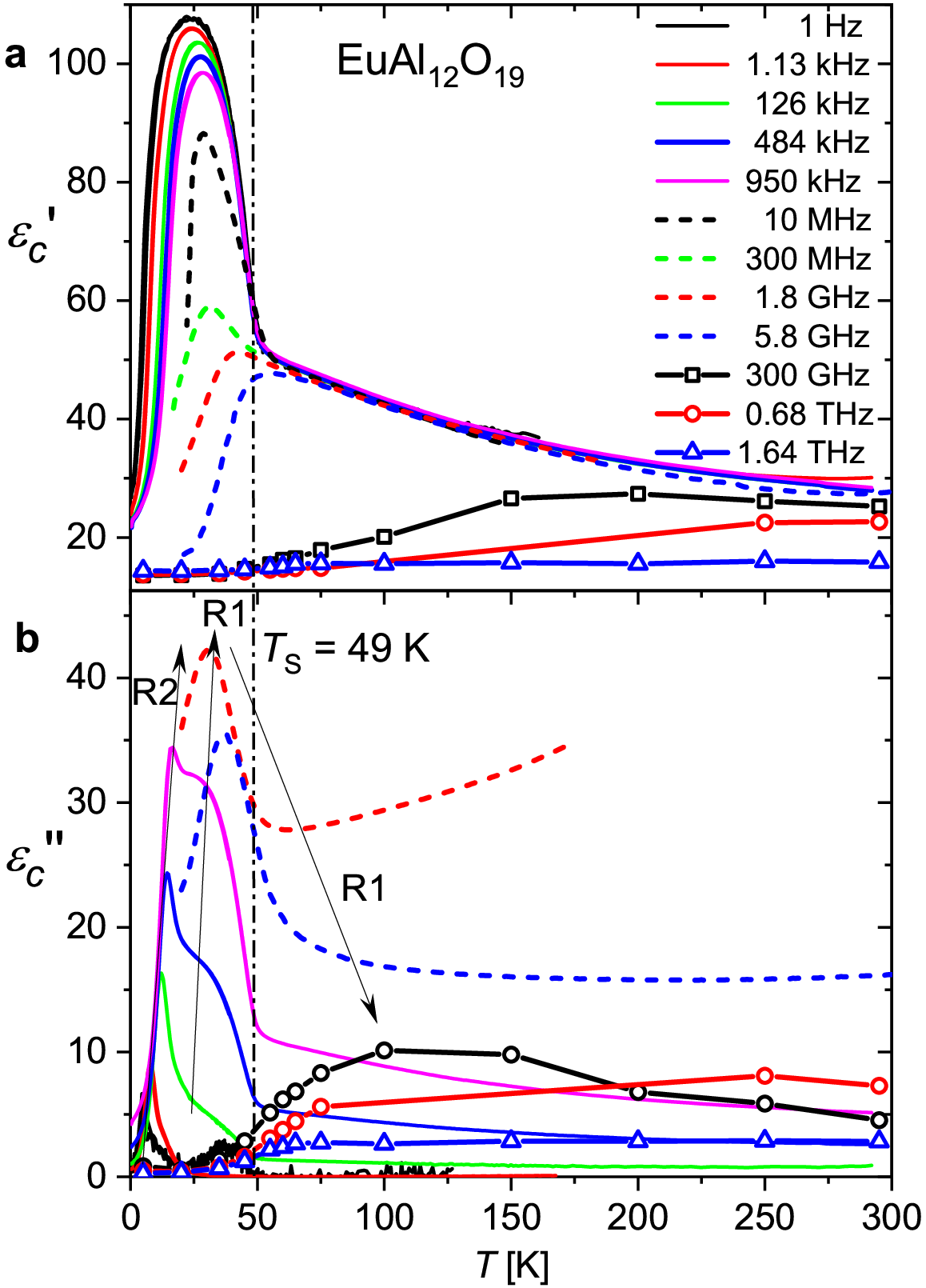}
\end{center}
\end{minipage}
\hfill
\begin{minipage}{0.47\linewidth}
\begin{center}
\includegraphics[width=0.9\linewidth]{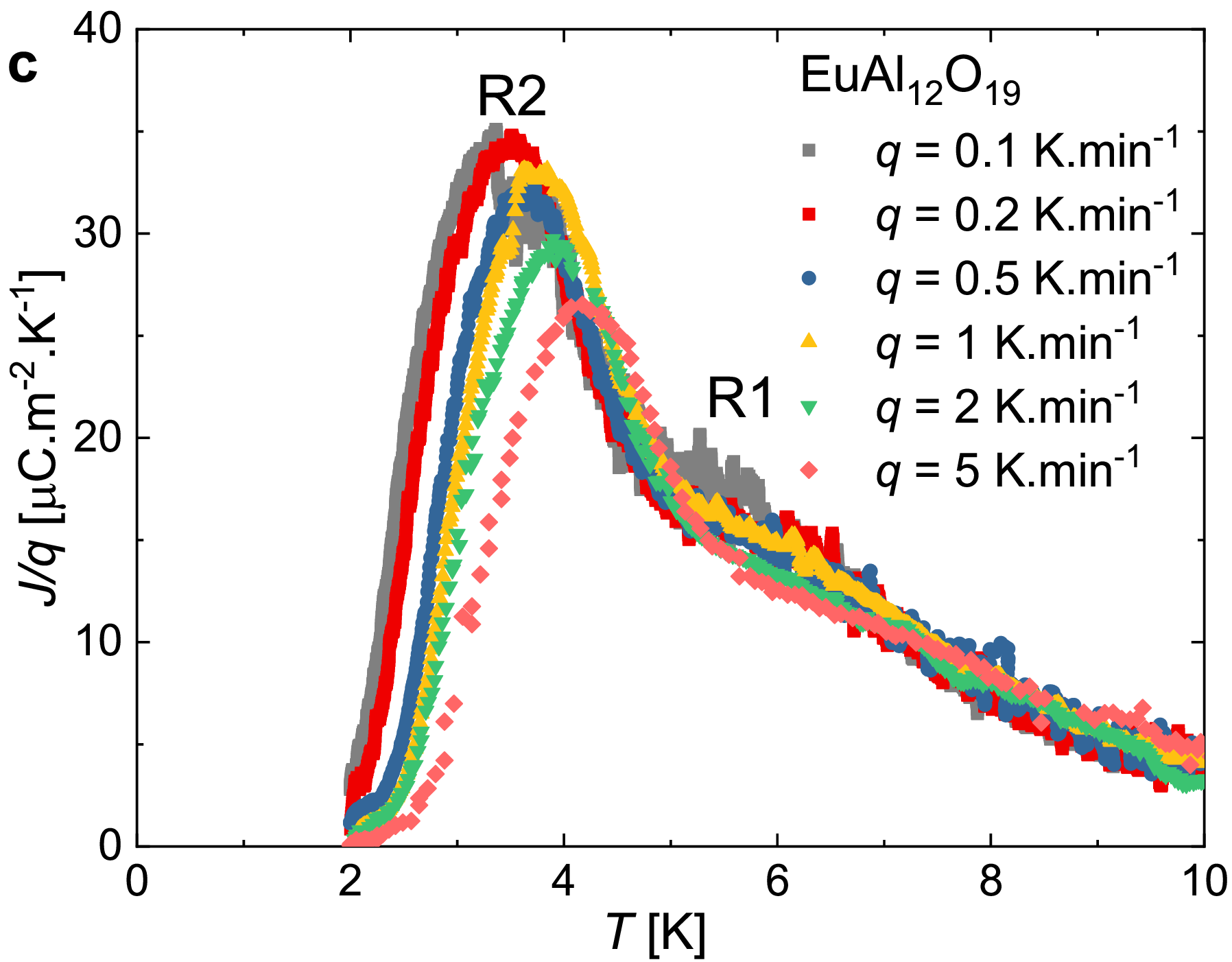}
\includegraphics[width=0.9\linewidth]{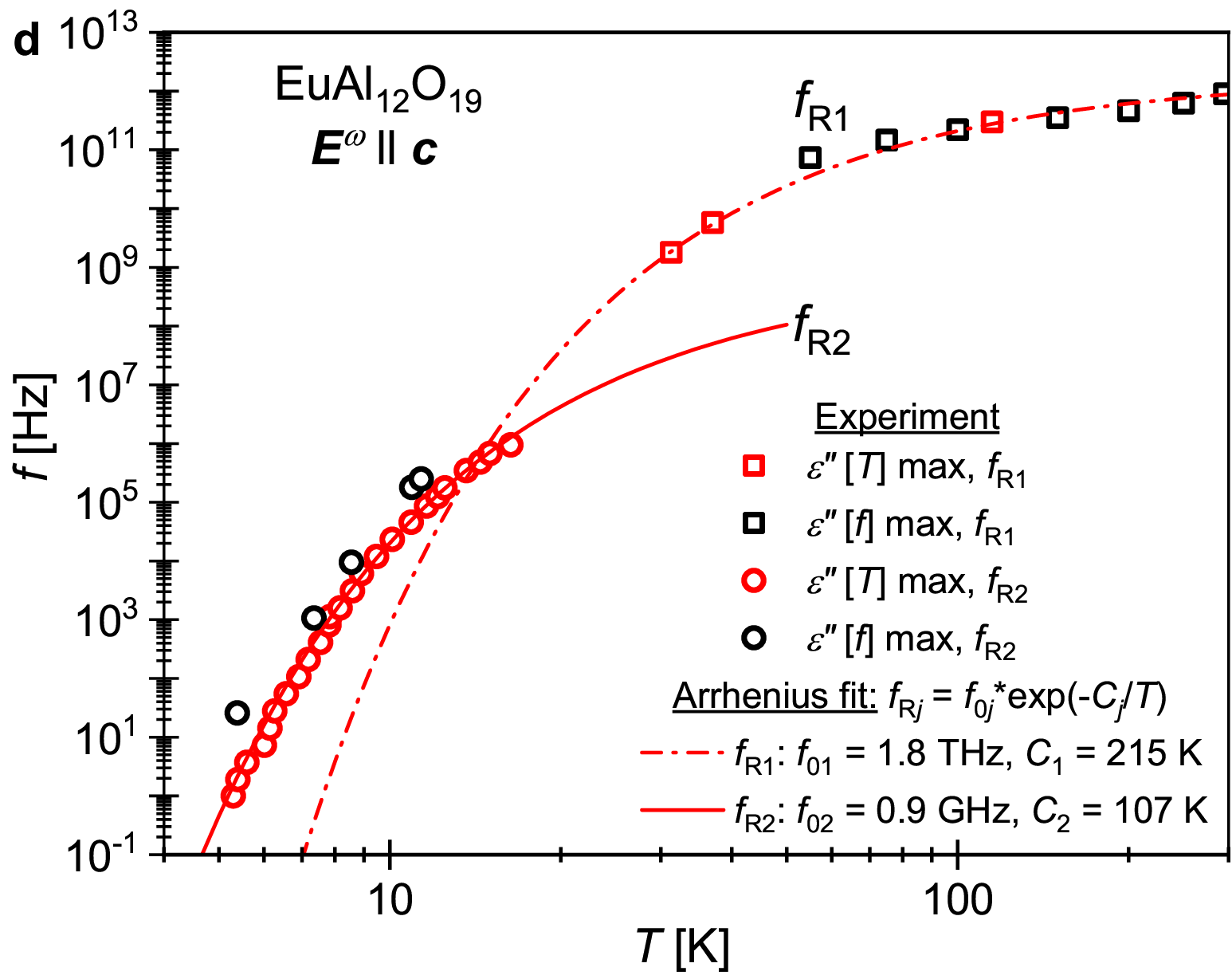}
\end{center}
\end{minipage}
\caption{Temperature dependence of \textbf{a} the real part $\varepsilon'_{\mathbf{c}}$ and \textbf{b} the imaginary part $\varepsilon''_{\mathbf{c}}$ of the dielectric permittivity at various frequencies. The relaxation R1 and R2 are indicated by a broad maximum and a sharp peak in $\varepsilon''_c (T)$. \textbf{c} Pyroelectric currents density divided by the heating rate $q$ for different heating rates. 
The crystal was cooled under an electric field of  $E=3.2$\,kV/cm along the $c$ axis.  \textbf{d} Temperature dependence of the $f_{\mathrm{R1}}$ and $f_{\mathrm{R2}}$ relaxation frequencies obtained as positions of maxima from either $\varepsilon''_{\mathbf{c}} (f)$ measured at a particular temperature or  $\varepsilon''_{\mathbf{c}} (T)$ spectra measured at a particular frequency. The lines are two Arrhenius fits. 
}
\label{disp}
\end{figure*}

We have investigated the dynamical properties of the electric dipoles by probing phonons and dielectric relaxations in EuAl$_{12}$O$_{19}$ over 13 orders of magnitude of frequencies from infrared (IR) frequencies ($\approx 10^{13}$\,Hz) down to 1\,Hz. The spectra of IR reflectivity for $\textbf{E}^{\omega}$$\parallel$~$\textbf{a}$ and $\textbf{E}^{\omega}$$\parallel$~$\textbf{c}$ reveal the appearance and reinforcement upon cooling of several IR active modes, without any sudden changes at $T_{\mathrm{S}}=49~$K (see \textbf{Figure~\ref{IR}(a-b)}). This evolution must be related to changes of local symmetry induced by the formation of short range antipolar correlations.  Nevertheless, the THz complex permittivity directly calculated from the measured complex transmittance spectra reveal a broad dielectric relaxation which slows down on cooling for $\textbf{E}^{\omega}$$\parallel$~$\textbf{c}$ (see \textbf{Figure~\ref{IR}(c-d)}; see also Figure~S8 in Supplemental for the behavior of the complex permittivity for $\textbf{E}^{\omega}$$\parallel$~$\textbf{a}$). The excitation behaves like an overdamped oscillator, resembling the behavior of a dielectric relaxation~\cite{Kamba2021}, and therefore we call it relaxation R1. The softening of its frequency $f_{\mathrm{R1}}$ quantitatively explains the C-W behavior of $\varepsilon'$ below $f=1.8\,$GHz because the product $\Delta \varepsilon_R f_{\mathrm{R1}}$  should be temperature independent (as $\Delta \varepsilon_R$ marks the contribution of R1 to $\varepsilon'$) and because the sum of phonon contributions in $\varepsilon'$ (obtained from the fits of IR reflectivity using a Lorentz three-parameter oscillator model)~\cite{Kamba2021} slightly decreases on cooling due to hardening of the TO$_{c}^{\mathrm{HT}}$ mode (Figure~\ref{IR}b). We propose that the microscopic origin of the R1 relaxation is connected to the dynamical disorder of the Al(5) cations within the AlO$_5$ bipyramids.

To follow the softening of the relaxation R1 upon cooling, we extended the measurement of complex permittivity down to the microwave (MHz-GHz) and low frequency (Hz-kHz) range (\textbf{Figure~\ref{disp}a-b}). The 
frequency of the relaxation R1 corresponds to the frequency where the maximum of $\varepsilon''_c(f)$ occurs.  Its softening is remarkable, since $f_{\mathrm{R1}}$ steadily decreases from $\approx$ 0.9\,THz 
at 295\,K to about 1\,kHz at 10\,K. In the Hz-kHz range $\varepsilon''_{\mathbf{c}}(T)$  exhibits a sharp peak ascribed to a second excitation of the relaxational type R2 and the maximum of the broad bump associated with the relaxation R1 is not fully discernible (Figure~\ref{disp}b). 
The R2 relaxation appears only below the transition $T_{\mathrm{S}}=49$\,K 
explaining the sudden change of slope of  $\varepsilon'_c(T)$ at this transition for frequencies below $f \approx$ 300\,MHz. It indicates that this transition opens a new route for the flipping of the electric dipoles. These dielectric permittivity measurements were complemented by  measurements of the elastic constant $C_{11}$ and $C_{33}$ from sound velocity experiments (See Figure~S9 in Supplemental). They also reveal an anomalous hardening along the $c$ axis below the phase transition at $T_{\mathrm{S}}=49$\,K.

The temperature dependence of the $f_{\mathrm{R1}}$  frequencies in the GHz-THz range and of the $f_{\mathrm{R2}}$ frequencies in the Hz-MHz range are well fitted by Arrhenius laws (\textbf{Figure~\ref{disp}d}) with the activation energies $C_1=215$\,K and $C_2=107$\,K for R1 and R2, respectively. Note that the Vogel-Fulcher law describing freezing of a dielectric relaxation (for example in relaxor ferroelectrics and dipolar glasses)~\cite{Courtens1984, Vugmeister1990, Hoechli1990, Kamba2021} could not better describe the data. In addition, attempts to model the data with the generalized Cochran formula $f_{\mathrm{R1}} = A(T-T^*)$ describing slowing-down of critical relaxation approaching a critical temperature $T^*$  could also not appropriately fit our data and therefore cannot explain the phase transition at $T_\mathrm{S}$. The Arrhenius behavior confirms that the dynamics of the electric dipoles is mainly thermally induced~\cite{Petzelt2021} and it implies an asymptotic freezing at $T=0\,K$, as expected for a classical liquid of either spins or electric dipoles~\cite{Harris1997, Moessner1998, Balents2010}.

The R1 and R2 relaxations can also help us to understand the unusual shape of the temperature dependence of the pyroelectric currents below 10\,K. Indeed, the pyroelectric current measurements show a peak around 4\,K shifting to higher temperature upon increasing the heating rate (\textbf{Figure~\ref{disp}c}). This rate-dependent shift confirms that it corresponds to a dynamical release of polarization and not to a phase transition~\cite{Vanderschueren2005, Jachalke2017}. 
The peak of polarization current is composed of a main peak around 4\,K and a broad shoulder around 6\,K, which can be ascribed to R2 and R1 relaxations, respectively. Thus, the measurements of pyroelectric currents allow us to follow qualitatively the relaxation frequencies $f_{\mathrm{R1}}$ and $f_{\mathrm{R2}}$  up to longer time scales given by heating rate ($t \approx 10 - 100$\,s), analogous to the use of thermoremanent magnetization measurements of slow dynamics in frustrated magnets
~\cite{Dupuis2002, Mamiya2014}.

\subsection{Synthesis of the results and interpretation}

Altogether our measurements enable us to draw the general picture of the behaviour of the lattice of electric dipoles in EuAl$_{12}$O$_{19}$. Around room temperature, the Al(5) jump between two potential minima in the AlO$_5$ bipyramids due to thermal fluctuations described by the relaxation R1. Upon cooling antipolar correlations between neighboring electric dipoles get reinforced and the classical electric dipole liquid forms. The R2 relaxation appearing below $T_\mathrm{S}=49$\,K can be ascribed to correlated flipping of electric dipoles. Both relaxation R1 and R2 soften upon cooling such that their characteristic time $1/f_\mathrm{R1}$ and $1/f_\mathrm{R2}$ diverge towards $T=0\,$K implying a nearly static disorder at low temperature. Following theoretical description of the excitations in the ITLAFM model~\cite{Jiang2005, Zhou2022}, the relaxation R2 can be attributed to the displacement of topologically protected objects such as topological strings and electric equivalent of spinons~\cite{Jiang2005, Zhou2022}. Compared to spin ices and ices of electric dipoles, where classical spin liquid states and their electric analogues were previously reported~\cite{Harris1997,Ramirez1999, Balents2010, Seshadri2006, McQueen2008, Coates2021}, the highly degenerate ground state of the ITLAFM and ITLAFE includes states with finite magnetization and polarization, respectively. This implies the possibility to record information locally by cooling under applied electric field~\cite{Hwang2008} as demonstrated in our pyroelectric current measurements (Figure~\ref{P}c and S1).

In real materials, we expect the degeneracy of the ground state of the ITLAFM and ITLAFE to be lifted by second nearest neighbor interactions. Together with interplane interactions they favor the formation of a long range three-dimensional order, implying a collapse of the entropy and the subsequent fulfilment of the third law of thermodynamics~\cite{Takasaki1986, Wang2014, Bradley2019}. Especially, the classical Monte Carlo simulations performed by Wang \textit{et al.}~\cite{Wang2014}  predicted the occurrence of a phase transition at $T=3$\,K for BaFe$_{12}$O$_{19}$, where the electric dipole liquid state would crystallize into the stripe antiferroelectric order (depicted in Figure~\ref{Fig1}f). 
The true ground state of EuAl$_{12}$O$_{19}$ may also be the antiferroelectric stripe order due to a lifting of the high degeneracy of the electric dipole liquid state by second nearest neighbor dipolar interactions. Since dielectric permittivity and pyro-electric currents measurements do not show any signature of antiferroelectric transition down to $0.3\,$K, the formation of the long-range antiferroelectric order is certainly avoided by a form of kinetic arrest, the  divergence of the timescale of the dynamics of the electric dipoles towards $T=0$\,K. 

\section{Conclusion}

The study of the dielectric and structural properties of the compound EuAl$_{12}$O$_{19}$ leads to the discovery of a frustrated antipolar phase. The frustration arises from a triangular lattice of uniaxial electric dipoles coupled by interactions favoring antipolar ordering of nearest neighbors. The electric dipoles form short-range correlations upon cooling, they remain disordered at any temperature and they do not show any conventional freezing at finite temperature. This behavior is similar to the one of the predicted classical spin liquid state in Ising triangular lattice antiferromagnets (ITLAFM). However, in contrast to the ITLAFM model, EuAl$_{12}$O$_{19}$ harbors a second order phase transition at $T_\mathrm{S}=49$\,K showing additional complexity in the electric analogue EuAl$_{12}$O$_{19}$. Further investigations of the nature of this transition will be needed for a better understanding of the electric dipole frustration.

The discovery of this unusual dielectric phase will motivate further investigations of the dielectric properties of the broad variety of compounds harboring the magnetoplumbite crystal structure~\cite{Li2016, Verstegen1974, Kimura1990, Venkateshwaran1999, Ni2022}. It will also promote further work on  electric dipole frustration in the pyrochlore structure~\cite{Seshadri2006, McQueen2008, Coates2021, HickoxYoung2022} and it will motivate the search for electric analogues of other class of geometrically frustrated magnets such as Kagome magnet and Kitaev magnets~\cite{Balents2010}. The classical electric dipole liquid state revealed in EuAl$_{12}$O$_{19}$ could be a good precursor for the formation of its quantum analogue, the quantum electric dipole liquid state, which could be valuable for quantum computing due to the presence of long-range quantum entanglement~\cite{Hotta2010, Hassan2018, Shen2016}.

\section{Experimental section}

\textbf{Density functional theory (DFT) calculations}
DFT+U calculations employed the full-potential linear augmented plane wave (FP-LAPW) method in the band structure program ELK~\cite{elk_code}, combined
with the generalized gradient approximation (GGA) parameterized by
Perdew-Burke-Ernzerhof~\cite{Perdew1996}. Our calculations have used the Hubbard $U=$ 2.7\,eV applied to Eu 4$f$ orbitals which was added to the Hamiltonian within the fully localized limit with double counting treatment~\cite{Anisimov1991}. The full Brillouin
zone has been sampled by 50 $k$-points, which is found to be
sufficient for system with such large unit cell. The maximum length
of the reciprocal lattice vectors for expanding the interstitial density
and potential has been enhanced to $20$ to correctly treat the situation
with a large mismatch between different atom muffin-tin radii. The crystal structure obtained from x-ray diffraction was relaxed prior to phonon calculations with and without maintaining the Al(5) in the $z=1/4$ plane. Phonon dynamical matrix elements were obtained from forces calculated self-consistently in supercells generated by the code PHONOPY~\cite{Togo2023}.

\textbf{Crystal growth and preparation}
Single crystals of EuAl$_{12}$O$_{19}$ were grown by the optical floating zone method from sintered polycrystalline rod. The polycrystalline rod was prepared from Al$_2$O$_3$ and Eu$_2$O$_3$ binary oxides, intimately mixed by grinding. The homogeneous mixture was packed into a latex mold under vibration, followed by hydrostatic pressing (2 tonnes for 15 minutes) to increase density. The pressed rods, of approximately 6\,mm diameter, were then sintered in air at 1473\,K for 24\,hrs. Growth was achieved in an optical floating zone furnace (FZ-T- 4000-VI-VPM-PC), under a flowing atmosphere of 5\%\,H$_2$ in Ar (0.2\,l/min), with counter rotation of upper and lower rods (50\,rpm). After slowly heating to the melting point, and stabilising the molten zone, the crystal was grown at a constant pulling speed of 3\,mm/h, with minor adjustments to furnace power during the growth to maintain a stable molten zone. We successfully obtained a single grain of about 5\,mm$\times$40\,mm$\times$3\,mm. This crystal was oriented by backscattering Laue diffraction. It was cleaved along the easy cleavage plane $ab$ or cut by a wire saw in other directions.

\textbf{Specific heat} Specific heat measurements were performed by the relaxation method using the PPMS from Quantum Design both with and without the $^3$He insert.

\textbf{Dielectric permitivitty} The low frequency (1\,Hz -- 1\,MHz) dielectric response was measured on a parallel plate (with a thickness of 305\,$\mu$m and an area of about 28\,mm$^2$) sample with gold electrodes using a Novocontrol Alpha-AN high performance impedance analyzer. A helium cryostat equipped with a $^3$He refrigerator was used to measure from room temperature down to 0.4\,K). The absolute value of the Hz-MHz $\varepsilon''_c$ may include a slowly varying background in the whole temperature range caused by rather long supply cables needed for experiments at very low temperatures. Measurements under applied electric fields were performed with the same setup but only down to 8\,K using Janis cryostat.

The dielectric measurement in the high-frequency range (1\,MHz – 1.8\,GHz) was carried out with an Agilent 4291B impedance analyser, a Novocontrol BDS 2100 coaxial sample cell and a  a Janis closed-cycle He cryostat. The sample was polished to a cylinder (height 2.2\,mm, diameter 1.92\,mm) with the crystallographic $c$ axis along the main axis, and with gold electrodes sputtered on the bases. Complex permittivity was calculated from measured complex impedance.

The microwave response at 5.8\,GHz was measured using the composite dielectric resonator method.\cite{Bovtun2008,Bovtun2011}. The TE$_{01\mathrm{\delta}}$ resonance frequency, quality factor and insertion loss of the base cylindrical dielectric resonator with and without the sample were recorded during heating from 10 to 400\,K with a temperature rate of 0.5\,K/min in a Janis closed-cycle He cryostat. We used a sample ($6.6\times5.2$\,mm plate, 2\,mm thick) without electrodes. Both $a$ and $c$ axes were in the plane. The sample was placed on top of the base dielectric resonator. The resonators were measured in the cylindrical shielding cavity using the transmission setup with a weak coupling by an Agilent E8364B network analyser. The complex permittivity of the sample corresponds to in-plane average and was calculated from the acquired resonance frequencies and quality factors of the base and composite resonators.

\textbf{Second harmonic generation (SHG)} The temperature dependence of the SHG signal was obtained using an optical setup powered by a Ti:sapphire femtosecond laser amplifier (Spitfire ACE) producing 40\,fs long pulses with a center wavelength of 800\,nm and repetition rate of 5\,kHz. The sample was placed in a continuous-He-flow cryostat Oxford Instruments with glass windows and illuminated by a collimated polarized beam with a diameter of 1.5–2.0\,mm and a pulse fluence up to 5\,mJ/cm$^2$ to verify the absence of SHG in the low-temperature phase. The resulting signal generated in a transmission configuration at 400\,nm was spectrally filtered, detected by an avalanche photodiode, and amplified by a lock-in amplifier (MFLI, Z\"urich Instruments), which ensures a background noise suppression. Schema of the SHG experimental setup is available in supplemental material of Ref.~\cite{Goian2023}. Independent transmission measurements in the optical region confirmed that the sample is transparent up to 350\,nm, thus the expected SHG signal at 400\,nm should be well measurable.

\textbf{Pyroelectric current measurements} The pyroelectric currents were measured on a (001) oriented crystal with a thickness of 305\,$\upmu$m and an area of about 28\,mm$^2$ with gold electrodes. The  crystal was cooled under electric field applied along the $c$ axis down to $T = 1.8~$K. Then the electric field was removed and after a dwelling time of 15 minutes, the pyrroelectric currents were measured upon heating. The PPMS from Quantum Design was used to cool and heat the sample with stable rates 0.1-5.0\,K/min. The electrical current was measured by a Keithley 617 electrometer. An homemade insert was used to apply voltage up to 180\,V on the crystal.

\textbf{Single crystal x-ray diffraction} Single-crystal X-ray diffraction was performed at the I19 beamline (Hutch 1) of the Diamond Light Source synchrotron in Harwell Oxford, UK~\cite{Allan2017}. The wavelength used is the Zr edge, specifically 17.9976\,keV ($\lambda$ = 0.68890\,\AA). We used the 3-circle diffractometer equipped with a Pilatus 2M detector. Cooling of the sample was carried out using the Oxford Cryosystems N-Helix. A crystal with approximate dimensions 60 × 60 × 10 $\upmu$m$^3$ was selected for the temperature-dependent measurement. Diffraction data were collected at 35, 80, 120 and 225\,K. Data processing (peak hunting, indexing, scaling, absorption correction) was performed using the Xia2-DIALS pipeline~\cite{Winter2010, Winter2018, BeilstenEdmands2020}. The output reflection file was used for structure solution and refinement in Jana2020~\cite{Petricek2023}, utilizing the charge flipping method via the built-in software SuperFlip~\cite{Palatinus2007}. At all temperatures, extinction conditions indicated the $P6_3/mmc$ space-group as best match. The phased structure factor information, output from Jana2020 after solution in the $P6_3/mmc$ space-group, was used to calculate the electron density distribution by maximum entropy method (MEM) via the Dysnomia software suite~\cite{Momma2013}. A map splitting the unit cell into 72×72×288 voxels, with a flat starting electron density distribution was used. No observable difference was seen when starting from a density distribution model based on the solved crystal structure. The presented crystal structure and electron density map visuals were created using VESTA~\cite{Momma2008}. The temperature dependence of the cell parameters (Figure~S7 in Supplemental) was measured in a $\theta-\theta$ diffractometer Siemens D500 equipped by a linear detector Mythen 1\,K and a Cu x-ray tube with Cu-K$\alpha_{1,2}$ radiation. The reciprocal space mapping of selected diffraction maxima (0 0 26) and (6 0 0) were used. The sample was cooled down to 3\,K inside a cryostat (ColdEdge) equipped with a piezorotator allowing the alignment of the sample in $\phi$ direction.

\textbf{Thermal expansion} High-resolution length changes along the $a$ and $c$ directions were measured using a miniature capacitance dilatometer~\cite{Rotter1998} connected to the AH2500A capacitance bridge implemented in PPMS from Quantum Design.

\textbf{Infrared (IR) spectroscopy} Low-temperature IR reflectance measurements in the frequency range 30$-$ 670\,cm$^{-1}$ ($1 – 20$\,THz) were performed using a Bruker IFS-113v Fourier-transform IR spectrometer equipped with a liquid-He-cooled Si bolometer (1.6\,K) serving as a detector. IR measurements were performed on the polished, over 500\,$\mu$m thick, plate-like sample with crystallographic axes $a$ and $c$ oriented parallel to plate edges.  Thus prepared sample was attached to a $4$-mm aperture and oriented in the desired way with respect to the polarized IR radiation. A continuous-He-flow cryostat Optistat with polyethylene windows was used for temperature control. 

\textbf{Terahertz spectroscopy} Temperature-dependent THz spectra of complex transmittance between 5 and 85\,cm$^{-1}$ were obtained using a custom-made time-domain spectrometer utilizing a Ti:sapphire femtosecond laser. Complex dielectric spectra were calculated directly from the complex transmittance spectra. The measurements were performed on a polished, free-standing, planeparallel, 509\,$\upmu$m thick, plate-like sample with crystallographic axes $a$ and $c$ oriented parallel to plate edges. The sample was attached to a $2.5$-mm aperture and oriented in the desired way with respect to the polarized THz radiation. A continuous-He-flow cryostat Oxford Instruments with mylar windows was used for temperature control. Details of the THz spectrometer and the principle of time-domain THz measurement are described elsewhere~\cite{Kuzel2010}.

\textbf{Sound velocity}: Sound velocity change measurements were performed using the ultrasound option for PPMS from Quantum Design (data in Figure~S9 in Supplemental). The measurements were performed using the phase comparison method~\cite{Luethi2007}. LiNbO$_{3}$ transducers were glued on Thiokol LP032 glue.

\medskip
\textbf{Supporting Information} \par 
Supporting Information is available from the Wiley Online Library.

\medskip
\textbf{Acknowledgements} \par 
We acknowledge funding from the Primus research program of the Charles University in Prague (Project number PRIMUS/22/SCI/016), from the Grant Agency of the Czech Technical University in Prague (Project No. SGS22/182/OHK4/3T/14), from Czech Science Foundation (Project No. 24-10791S) and from the European Union and the Czech Ministry of Education, Youth and Sports (Project TERAFIT-CZ.02.01.01/00/22\_008/0004594). Crystal growth and some of the measurements of physical properties were carried out in the MGML (http://mgml.eu/), which is supported within the program of Czech Research Infrastructures (project no. LM2023065). We acknowledge the Diamond Light Source, I19 (proposal CY33159), for the allocation of beam time and the financing of travels to their laboratory. 
\medskip

%
\bibliographystyle{MSP}
\bibliography{EuAl12O19-ferroelectricity}

\end{document}


\title{A frustrated antipolar phase analogous to classical spin liquids 
\\\textbf{Supplemental information}}

\author{G. Bastien}
\affiliation{Charles University, Faculty of Mathematics and Physics, Department of Condensed Matter Physics, Ke Karlovu 5, 121 16 Prague 2, Czech Republic}
\author{D. Rep\v{c}ek}
\affiliation{Institute of Physics, Czech Academy of Sciences, Na Slovance 2, 182 00 Prague, Czech Republic}
\affiliation{Czech Technical University in Prague, Faculty of Nuclear Sciences and Physical Engineering, Department of Solid State Engineering, B\v{r}ehová 7, 115 19 Prague 1, Czech Republic}
\author{A. Eli\'a\v{s}}
\affiliation{Charles University, Faculty of Mathematics and Physics, Department of Condensed Matter Physics, Ke Karlovu 5, 121 16 Prague 2, Czech Republic}
\author{A. Kancko}
\affiliation{Charles University, Faculty of Mathematics and Physics, Department of Condensed Matter Physics, Ke Karlovu 5, 121 16 Prague 2, Czech Republic}
\author{Q. Courtade}
\affiliation{Charles University, Faculty of Mathematics and Physics, Department of Condensed Matter Physics, Ke Karlovu 5, 121 16 Prague 2, Czech Republic}
\author{T. Haidamak}
\affiliation{Charles University, Faculty of Mathematics and Physics, Department of Condensed Matter Physics, Ke Karlovu 5, 121 16 Prague 2, Czech Republic}
\author{M. Savinov}
\affiliation{Institute of Physics, Czech Academy of Sciences, Na Slovance 2, 182 00 Prague, Czech Republic}
\author{V. Bovtun}
\affiliation{Institute of Physics, Czech Academy of Sciences, Na Slovance 2, 182 00 Prague, Czech Republic}
\author{M. Kempa}
\affiliation{Institute of Physics, Czech Academy of Sciences, Na Slovance 2, 182 00 Prague, Czech Republic}
\author{K. Carva}
\affiliation{Charles University, Faculty of Mathematics and Physics, Department of Condensed Matter Physics, Ke Karlovu 5, 121 16 Prague 2, Czech Republic}
\author{M. Vali\v{s}ka}
\affiliation{Charles University, Faculty of Mathematics and Physics, Department of Condensed Matter Physics, Ke Karlovu 5, 121 16 Prague 2, Czech Republic}
\author{P. Dole\v{z}al}
\affiliation{Charles University, Faculty of Mathematics and Physics, Department of Condensed Matter Physics, Ke Karlovu 5, 121 16 Prague 2, Czech Republic}
\author{M. Kratochv\'ilov\'a}
\affiliation{Charles University, Faculty of Mathematics and Physics, Department of Condensed Matter Physics, Ke Karlovu 5, 121 16 Prague 2, Czech Republic}
\author{S. A. Barnett}
\affiliation{ Diamond Light Source, Chilton, Didcot, Oxfordshire, OX11 0DE, United Kingdom}
\author{P. Proschek}
\affiliation{Charles University, Faculty of Mathematics and Physics, Department of Condensed Matter Physics, Ke Karlovu 5, 121 16 Prague 2, Czech Republic}
\author{J. Prokle\v{s}ka}
\affiliation{Charles University, Faculty of Mathematics and Physics, Department of Condensed Matter Physics, Ke Karlovu 5, 121 16 Prague 2, Czech Republic}
\author{C. Kadlec}
\affiliation{Institute of Physics, Czech Academy of Sciences, Na Slovance 2, 182 00 Prague, Czech Republic}
\author{P. Ku\v{z}el}
\affiliation{Institute of Physics, Czech Academy of Sciences, Na Slovance 2, 182 00 Prague, Czech Republic}
\author{R. H. Colman}
\affiliation{Charles University, Faculty of Mathematics and Physics, Department of Condensed Matter Physics, Ke Karlovu 5, 121 16 Prague 2, Czech Republic}
\author{S. Kamba}
\affiliation{Institute of Physics, Czech Academy of Sciences, Na Slovance 2, 182 00 Prague, Czech Republic}

\maketitle

\section{Crystal structure relaxation and Phonon spectra computation}

\begin{figure}[h]
\renewcommand{\figurename}{Figure S}
\includegraphics[width=0.9\linewidth]{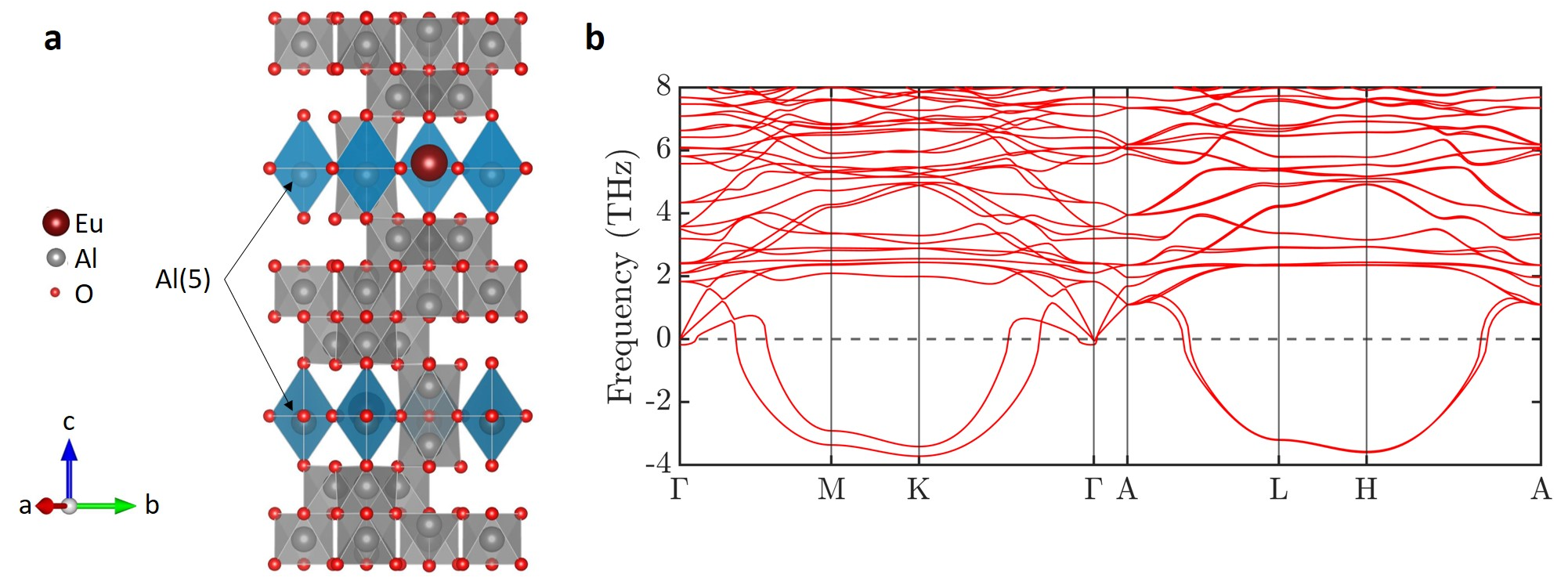}
\caption{\textbf{a} Schematic view of the ferroelectric  crystal structure obtained upon relaxation from the experimentally determined crystal structure of EuAl$_{12}$O$_{19}$. The relaxation was performed on a single unit cell with periodic boundary condition. In the relaxed structure, the Al(5) ions are displaced downwards from the centre of their bipyramid. \textbf{b} Phonon dispersion structure calculated for the relaxed structure. Only phonons below 8\,THz are shown to focus on the low energy branches and soft modes. Imaginary frequencies are represented by negative values.}
\label{FigS1}
\end{figure}

Since DFT+U calculations performed with the magnetoplumbite-type
centrosymmetric structure (space group P$6_{3}/mmc$) revealed its instability (see Fig.~1g in the main text), we performed a second calculation with an off-centering of Al(5) in the bipyramid AlO$_{5}$. The unit cell of the crystal structure of EuAl$_{12}$O$_{19}$ contains two Al(5) located in two different triangular layers of electric dipoles (Fig.~1c in the main text or Fig S\ref{MEMsup}). We have performed structural relaxation both with parallel and antiparallel alignment of these two electric dipoles starting from the structure
determined by x-ray diffraction, where only the Al(5) ion violates the symmetry by the mirror plane $z=1/4$. The centrosymmetric structure without electric dipoles was also relaxed for comparison. Total calculated energies of an infinite periodic system with both ordering and with an absence of electric dipoles  were compared and the ferroelectric order is the most favorable. This indicates the tendency to form electric dipoles, as well as preference for parallel alignment of dipoles located on adjacent triangular layers. A schematic view of the relaxed structure is shown in Fig.~S\ref{FigS1}a and the atomic coordinates obtained after relaxation are shown in Tab~S\ref{Tabrelax}. Upon relaxation, the Al(5) ions are displaced by $\delta=0.25\,\mathring{\mathrm{A}}$ in good agreement with the experimental value of $\delta=0.24\,\mathring{\mathrm{A}}$ from MEM analysis at $T=35\,$K. In addition, the relaxation indicates a vertical displacement of the O(1) ions forming the basal triangle of the bipyramid AlO$_5$  of $\delta_O=0.07\,\mathring{\mathrm{A}}$, which cannot be resolved from x-ray diffraction. The relaxed crystal structure belongs to the non-centrosymmetric space group P$6_{3}mc$.

As the next step we have evaluated phonon spectra for an infinite
periodic system with the unit cell deduced from the relaxation (Fig.~S\ref{FigS1}b). Two soft modes are still present, but reach imaginary frequencies (plotted as negative values) only near points M, K, L, H; not at $\Gamma$ and A.
This corresponds to $k$-points approaching the boundary of the Brillouin
zone in the planar direction. These two soft modes are dominated by
the displacement of Al in the bipyramid AlO$_{5}$ in the $c$ direction and their dispersion indicates that phonon modes with Al displaced oppositely in neighbouring cell in the $ab$ plane would be of lowest energy. These results hint for a preference for antiparrallel alignement of the nearest neighbors as it was further confirmed by the computation of energy in a system made of two unit cells (see main text, section 2.1). It implies the presence of antipolar interactions within the triangular lattice of electric dipoles in the $ab$ plane which is the key ingredient for the formation of geometrical frustration. 

\begin{table}[!h]
\renewcommand{\tablename}{Table S}
\caption{Atomic coordinates of the relaxed crystal structure of EuAl$_{12}$O$_{19}$. The cell parameters were kept constant at the experimental values obtained from structural refinement of single crystal diffraction at $T=35$\,K, $a$ = 5.56480 $\mathring{\mathrm{A}}$ and $c$ = 21.9912 $\mathring{\mathrm{A}}$. The left column gives the results of a relaxation with the inversion symmetry initially broken by slightly displacing Al(5) atom in the $z$ direction resulting in a ferroelectric crystal structure (space group $P6_3mc$). The right coloumn gives  the results of a relaxation performed upon starting with a centrosymmetric crystal structure (space group $P6_3/mmc$).}
 \label{Tabrelax} 
        \begin{tabular}{cccccccc}
		\hline
  \multicolumn{4}{c}{Al(5) off-center} & &\multicolumn{3}{c}{centrosymmetric structure}\\
        \hline
	            {Atom}            & {$x$}             & {$y$}             & {$z$}             &
                 &
             {$x$}             & {$y$}             & {$z$}  
                        \\  
                        \hline                {Eu} &
              {0.33333} &	{0.66667} & {0.75000}& &{0.33333} &	{0.66667} & {0.75000}\\
              {Al(1)} &
              {0.16873} &	{0.33747} & {0.60542}& ~~~~~~~~&
              0.16808	& 0.33617& 	0.60789\\
              {Al(2)} &
              {0.33333} &	{0.66667} & {0.18801} & & 0.33333	& 0.66667 & 0.19034 \\
              {Al(3)} &
              {0.00000} &	{0.00000} & {0.99768} & & {0.00000} &	{0.00000} & {0.00000}  \\
           		{Al(4)} &
              {0.33333} &	{0.66667} & {0.02675} & & 0.33333 &	0.66667 &	0.02909\\
              {Al(5)} & 
              {0.00000} &	{0.00000} & {0.23880}& & 0.00000	& 0.00000	& 0.25000\\
              {O(1)} &
              {0.18138} &	{0.36275} & {0.24690}& &0.18178	& 0.36356 &	0.25000\\
              {O(2)} &
              {0.50255} &	{0.00510} & {0.14571}& & 0.50213 &	0.00426 &	0.14774\\
              {O(3)} &
              {0.00000} &	{0.00000} & {0.14798}& & 0.00000	& 0.00000 &	0.14881\\
              {O(4)} &
              {0.15428} &	{0.30859} & {0.05023}& & 0.15465	& 0.30934 &	0.05169\\
              {O(5)} &
              {0.33333} &	{0.66667} & {0.55179} & & 0.33333 &	0.66667 &	0.55380\\
              \hline         
	\end{tabular}
\end{table}

\section{Electric field dependence of the pyroelectric currents}

\begin{figure}[h!]
\renewcommand{\figurename}{Figure S}
	\begin{center}
		\includegraphics[width=0.6\linewidth]{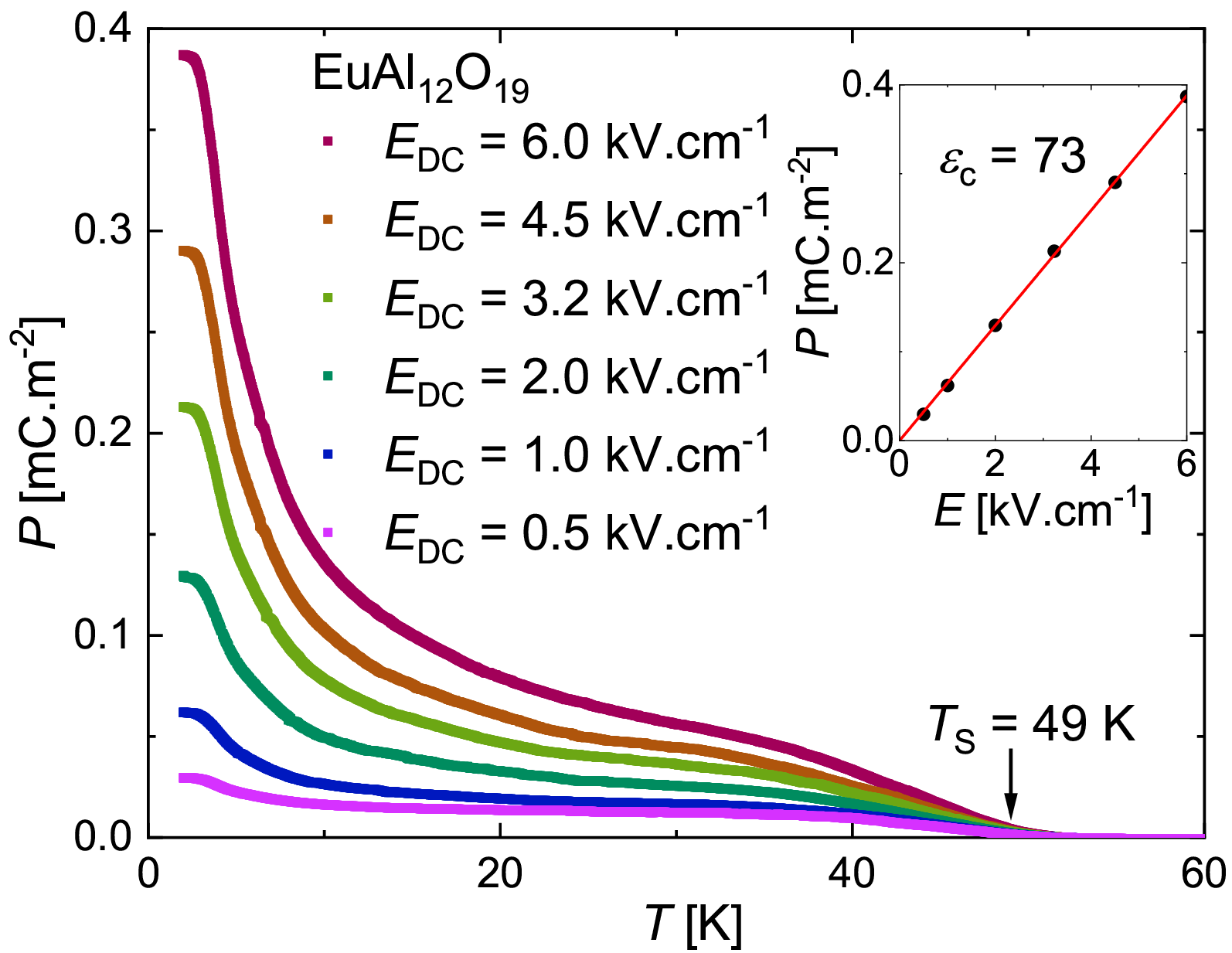}
		\caption{Temperature dependence of polarization along the $c$ axis calculated from pyrocurrent measurements as a function of poling electric field. The measurements were performed upon heating at a constant rate of $q=2\,$K/min in absence of external electric field. The inset shows the polarization at $T=2\,$K as a function of the poling field $E$. It confirms the assumption of a linearly induced polarization $P=\varepsilon_0 \varepsilon_c E$, with a static dielectric permittivity of $\varepsilon_c=73$.}
		\label{PE} 
	\end{center}
\end{figure}

Pyroelectric currents were measured using the common procedure for the study of ferroelectric materials~\cite{Jachalke2017}. A single crystal of EuAl$_{12}$O$_{19}$ was cooled under a bias electric field applied along the $c$ axis down to $T=2$\,K. Then the electric field was removed and after a dwelling time of 15\,minutes the pyroelectric current of the crystal was measured upon heating and the temperature dependence of the polarization was obtained by its integration. The polarization as a function of temperature and electric field is shown in Fig.~S\ref{PE}. The polarization at $T=2\,$K shows no signs of saturation (see inset of Fig.~S\ref{PE}), it is proportional to the electric field $P=\varepsilon_0 \varepsilon_c E$ with the value $\varepsilon_c$=73, which agrees rather well with the experimental value of the dielectric permittivity in Fig.~2b of the main text. Fig.~S\ref{PE} indicates that the polarization is induced by the applied electric field and it does not correspond to any spontaneous polarization. This electric field-induced polarization can be observed because the dynamics of the electric dipoles AlO$_5$ at $T=2\,$K are slow enough to maintain the polarization after the removal of the electric field at least on the timescale of an hour (see Fig5c-d). The relaxation of the polarization with time is evidenced by a significant decrease in polarization when heated (without field) from 2 to 20\,K.

We expect the polarization of the ITLAFE to deviate from linear behavior at finite electric field and to show a saturation at $P=n\mu/3$ analogous to magnetization plateaus in triangular magnets~\cite{Takasaki1986, Hwang2008}. This saturation would be related to the formation of an improper ferroelectric up-up-down order (see Fig.~1\textbf{f} in the main text). Using the  magnitude of the electric dipole $\mu=0.97\,e$.$\mathring{\mathrm{A}}$ extracted from the Curie-Weiss fit and the density of electric dipoles $n=3.38.10^{27}$\,m$^{-3}$, we can estimate the polarization of this phase at $P \approx 18$\,mC/m$^2$. This value exceeds by almost two orders of magnitude the polarization $P=0.39$\,mC/m$^2$ achieved at the highest electric field applied, $E=6\,$kV/cm. Thus the observed linearity of the polarization in applied electric field does not show any contradiction with the expectation of  a polarization plateau in the ITLAFE.

\section{Dielectric permittivity under applied electric bias field}
\begin{figure}[h!]
\renewcommand{\figurename}{Figure S}	
 \begin{center}
		\includegraphics[width=0.6\linewidth]{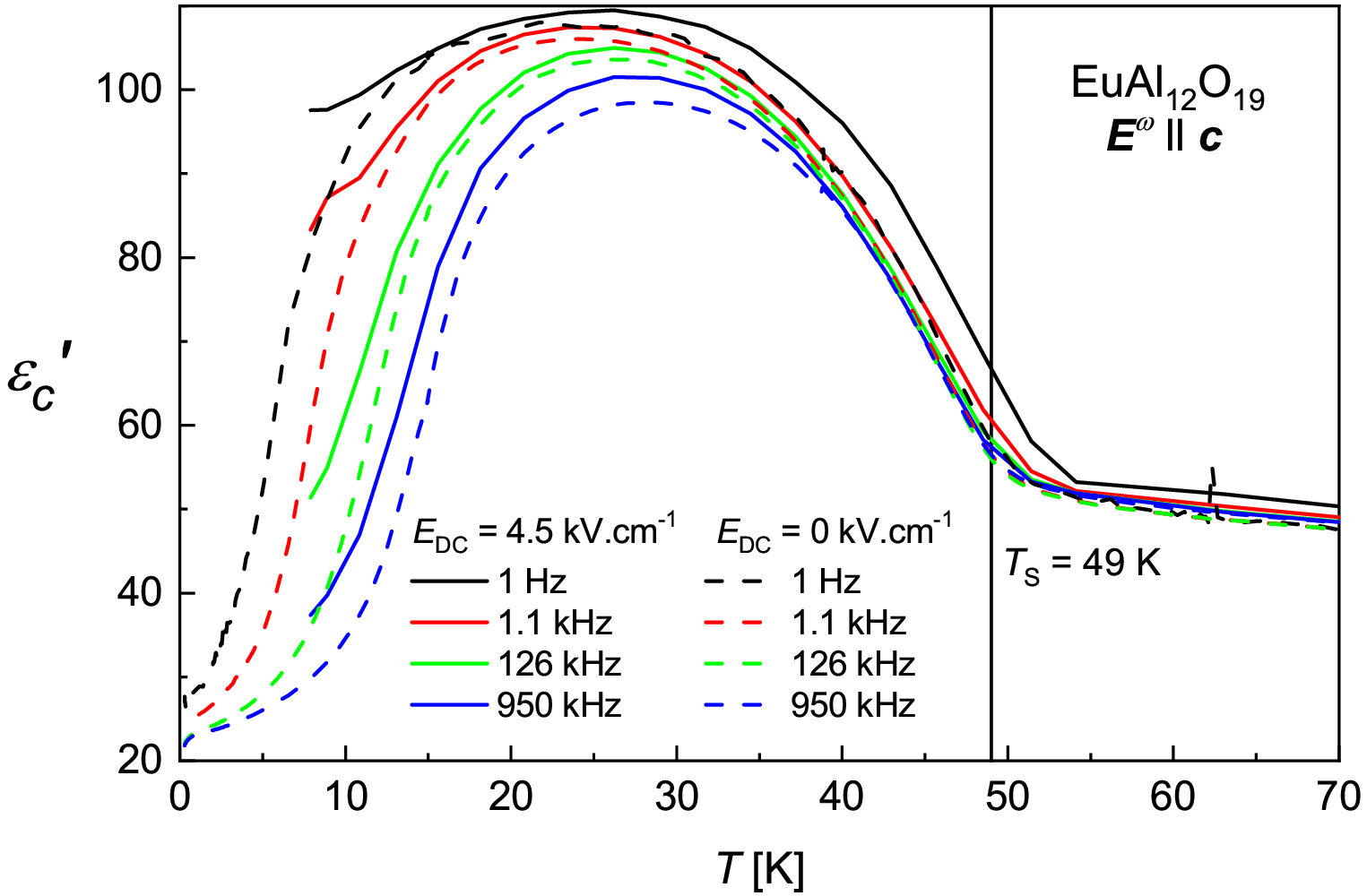}
		\caption{Comparison of temperature dependence of the dielectric permittivity measured under bias electric field $E_{\mathrm{DC}} = 4.5$\,kV/cm (applied along the $c$ axis) with the dielectric permittivity measured without electric field.}
		\label{fig:Hz-MHz_eps1_dispersion_with_electric_field} 
	\end{center}
\end{figure}

In ferroelectric states, the vibration of the domain walls can contribute strongly to the dielectric permittivity~\cite{Kumar2013, Kamba2017, Goian2023}. In addition, the number of domain walls is expected to decrease upon application of a static electric field $E_{\mathrm{DC}}$, consequently, the dielectric response should be diminished.\cite{Kumar2013, Kamba2017, Goian2023}. Thus measurements of the dielectric permittivity under applied static (DC) electric field $E_{\mathrm{DC}}$ are often used to distinguish between the contribution of domain walls and the intrinsic contribution from the bulk~\cite{Kumar2013, Kamba2017, Goian2023}.

However, our experiments with $E_{\mathrm{DC}} = 0$\,kV/cm and $E_{\mathrm{DC}} = 4.5$\,kV/cm showed no significant difference in $\varepsilon_c^{'}$ (see Fig.~S\ref{fig:Hz-MHz_eps1_dispersion_with_electric_field}). The observed differences are due to the different cryostats used and therefore lie within the accuracy of the measurements. The dielectric response of the relaxation R2 is thus practically the same and it cannot be ascribed to the contribution of ferroelectric domain walls. These results further support the absence of ferroelectric order.

\section{Absence of superstructure peaks in the single crystal x-ray diffraction data}

\begin{figure}[h!]
\renewcommand{\figurename}{Figure S}
	\begin{center}
		\includegraphics[width=0.8\linewidth]{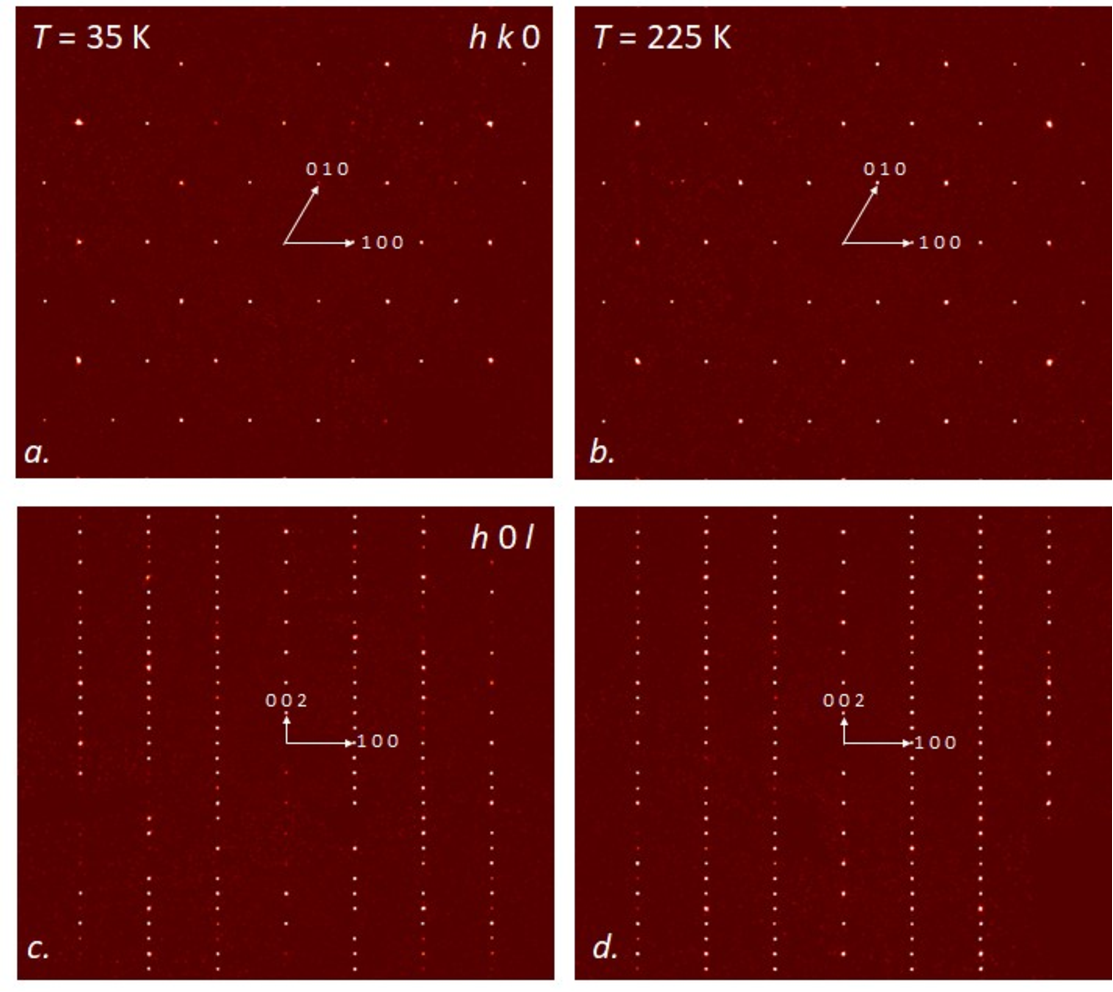}
		\caption{Reconstructed reciprocal space maps within the $hk0$ (a and b) and $h0l$ (c and d) planes at $T=35$\,K and $T=225$\,K, without absorption correction.}
		\label{maps} 
	\end{center}
\end{figure}

To further show the absence of superstructure in EuAl$_{12}$O$_{19}$ at $T_\mathrm{S}=49$\,K, we plotted reconstructed reciprocal space maps in Fig.~S\ref{maps} obtained at $T=35$\,K and $T=225$\,K. Long-range antiferroelectric order or improper ferroelectric order with antipolar alignment of dipoles located in the same $ab$ plane would imply a multiplication of the unit cell that would result in additional peaks within these reciprocal space maps. No superstructure peaks are observed at $T=35$\,K and their absence shows the absence of multiplication of the unit cell at $T_\mathrm{S}=49$\,K. This transition may still be associated with minor changes of the crystal structure beyond the resolution of the x-ray diffraction experiments, however the present measurements demonstrate that the electric dipoles AlO$_5$ do not form any long range order at this transition.

\newpage

\section{Refinement details of temperature dependent single crystal diffraction}

The detailed results from the refinement of the data from SCXRD are given in Tables ~S\ref{table:SCXRDtable35K} to~S\ref{table:SCXRDtable225K}.

\begin{table}[!h]
\renewcommand{\tablename}{Table S}
\caption{Details of the single crystal structure solution of EuAl$_{12}$O$_{19}$ using the I19 single crystal diffractometer at the Diamond Light Source, at $T=35$\,K}
\label{table:SCXRDtable35K}
        \begin{tabular}{cccccccccccccccccccccccccc}
		\hline
		
		\multicolumn{5}{c}{\begin{tabular}[c]{@{}c@{}}Space   group: \\    $P6_3/mmc$ (\#194, setting 1)\end{tabular}}                                                 & \multicolumn{5}{c}{$a$ = 5.56480(3) $\mathring{\mathrm{A}}$}                                                         & \multicolumn{5}{c}{$c$ = 21.9912(2) $\mathring{\mathrm{A}}$}                                                                                                          & \multicolumn{6}{c}{$V$ = 589.8 $\mathring{\mathrm{A}}^3$}                                                                                                               & \multicolumn{5}{c}{$Z$ = 2}                                                                                                                           \\
		
		\hline
		
		\multicolumn{4}{c}{\begin{tabular}[c]{@{}c@{}}Radiation:   \\    X-rays synchrotron \\    ($\lambda$ = 0.6889 $\mathring{\mathrm{A}}$)\end{tabular}} & \multicolumn{3}{c}{\begin{tabular}[c]{@{}c@{}}Reflections   \\ collected/unique/used: \\ 13099/725/588\\\end{tabular}} & \multicolumn{3}{c}{\begin{tabular}[c]{@{}c@{}}Final $R$   indices: \\    $R$ = 3.10 \% \\    $wR_2$ = 7.39 \% \\  Parameters: 42\end{tabular}} & \multicolumn{7}{c}{\begin{tabular}[c]{@{}c@{}}Maximum   difference:\\  Peaks \& Holes \\        0.56 $e\mathring{\mathrm{A}}^{-3}$   \&   $-$0.68 $e\mathring{\mathrm{A}}^{-3}$ \end{tabular}} & \multicolumn{4}{c}{\begin{tabular}[c]{@{}c@{}}Density: \\    $\rho$ = 4.3908 g.cm$^{-3}$\end{tabular}} & \multicolumn{4}{c}{\begin{tabular}[c]{@{}c@{}}  Absorption \\ coefficient: \\    $\mu$ = 5.826 mm$^{-1}$\end{tabular}} \\
		
		\hline
		
		\multirow{2}{*}{$T$ (K)}             & \multirow{2}{*}{Atom}            & \multirow{2}{*}{Site}            & \multicolumn{3}{c}{\multirow{2}{*}{$x$}}                                                           & \multicolumn{2}{c}{\multirow{2}{*}{$y$}}                                                 & \multirow{2}{*}{$z$}                                        & \multicolumn{2}{c}{\multirow{2}{*}{Occ.}}                   & \multicolumn{15}{c}{$U_{\mathrm{ij}}$ ($\times$10$^4  \mathring{\mathrm{A}}^2$)}                                                                                                                                                                                                                                             \\ \cline{12-25}
		&                                  &                                  & \multicolumn{3}{c}{}                                                                             & \multicolumn{2}{c}{}                                                                   &                                                           & \multicolumn{2}{c}{}                                        & \multicolumn{2}{c}{$U_{11}$}                    & \multicolumn{2}{c}{$U_{22}$}                 & \multicolumn{2}{c}{$U_{33}$}              & \multicolumn{3}{c}{$U_{12}$}                          & \multicolumn{2}{c}{$U_{13}$}       & \multicolumn{2}{c}{$U_{23}$}       & \multicolumn{1}{c}{$U_{\mathrm{eq}}$}            &               \\
		\hline
		
		\multirow{11}{*}{35 K}             & Eu                               & 2d                               & \multicolumn{3}{c}{$\frac{1}{3}$}                                                                          & \multicolumn{2}{c}{$\frac{2}{3}$}                                                                  & $\frac{3}{4}$                                                      & \multicolumn{2}{c}{1}                                       & 20(1)                  & \multicolumn{3}{c}{$U_{11}$}                                      & \multicolumn{3}{c}{25(2)}                           & \multicolumn{2}{c}{$\frac{1}{2}U_{11}$}    & \multicolumn{2}{c}{0}         & \multicolumn{2}{c}{0}         & 22(1)              &               \\
		& Al(1)                            & 12k                              & \multicolumn{3}{c}{0.16834(7)}                                                                   & \multicolumn{2}{c}{0.33668(7)}                                                         & 0.60833(4)                                                & \multicolumn{2}{c}{1}                                       & 14(3)                  & \multicolumn{3}{c}{$U_{11}$}                                      & \multicolumn{3}{c}{29(4)}                           & \multicolumn{2}{c}{6(3)}      & \multicolumn{2}{c}{$-U_{23}$}      & \multicolumn{2}{c}{2(1)}      & 19(3)              &               \\
		& Al(2)                            & 4f                               & \multicolumn{3}{c}{$\frac{1}{3}$}                                                                            & \multicolumn{2}{c}{$\frac{2}{3}$}                                                                  & 0.19073(7)                                                & \multicolumn{2}{c}{1}                                       & 17(4)                  & \multicolumn{3}{c}{$U_{11}$}                                      & \multicolumn{3}{c}{22(6)}                           & \multicolumn{2}{c}{$\frac{1}{2}U_{11}$}    & \multicolumn{2}{c}{0}         & \multicolumn{2}{c}{0}         & 18(3)              &               \\
		& Al(3)                            & 2a                               & \multicolumn{3}{c}{0}                                                                            & \multicolumn{2}{c}{0}                                                                  & 0                                                         & \multicolumn{2}{c}{1}                                       & 12(5)                  & \multicolumn{3}{c}{$U_{11}$}                                      & \multicolumn{3}{c}{8(7)}                            & \multicolumn{2}{c}{$\frac{1}{2}U_{11}$}    & \multicolumn{2}{c}{0}         & \multicolumn{2}{c}{0}         & 11(4)              &               \\
		& Al(4)                            & 4f                               & \multicolumn{3}{c}{$\frac{1}{3}$}                                                                            & \multicolumn{2}{c}{$\frac{2}{3}$}                                                                  & 0.02831(7)                                                & \multicolumn{2}{c}{1}                                       & 17(4)                  & \multicolumn{3}{c}{$U_{11}$}                                      & \multicolumn{3}{c}{21(6)}                           & \multicolumn{2}{c}{$\frac{1}{2}U_{11}$}    & \multicolumn{2}{c}{0}         & \multicolumn{2}{c}{0}         & 18(3)              &               \\
		& Al(5)                            & 4e                               & \multicolumn{3}{c}{0}                                                                            & \multicolumn{2}{c}{0}                                                                  & 0.24015(12)                                               & \multicolumn{2}{c}{$\frac{1}{2}$}                                       & 15(5)                  & \multicolumn{3}{c}{$U_{11}$}                                      & \multicolumn{3}{c}{30(2)}                           & \multicolumn{2}{c}{$\frac{1}{2}U_{11}$}    & \multicolumn{2}{c}{0}         & \multicolumn{2}{c}{0}         & 19(8)              &               \\
		& O(1)                             & 6h                               & \multicolumn{3}{c}{0.1816(2)}                                                                    & \multicolumn{2}{c}{0.3632(5)}                                                          & $\frac{1}{4}$                                                         & \multicolumn{2}{c}{1}                                       & 34(8)                  & \multicolumn{3}{c}{21(10)}                                   & \multicolumn{3}{c}{39(11)}                          & \multicolumn{2}{c}{$\frac{1}{2}U_{22}$}    & \multicolumn{2}{c}{0}         & \multicolumn{2}{c}{0}         & 33(7)              &               \\
		& O(2)                             & 12k                              & \multicolumn{3}{c}{0.5022(3)}                                                                    & \multicolumn{2}{c}{0.00440(16)}                                                        & 0.14809(9)                                                & \multicolumn{2}{c}{1}                                       & 19(7)                  & \multicolumn{3}{c}{18(5)}                                    & \multicolumn{3}{c}{26(7)}                           & \multicolumn{2}{c}{$\frac{1}{2}U_{11}$}    & \multicolumn{2}{c}{2$U_{23}$}      & \multicolumn{2}{c}{3(3)}      & 21(5)              &               \\
		& O(3)                             & 4e                               & \multicolumn{3}{c}{0}                                                                            & \multicolumn{2}{c}{0}                                                                  & 0.14856(14)                                               & \multicolumn{2}{c}{1}                                       & 12(7)                  & \multicolumn{3}{c}{$U_{11}$}                                      & \multicolumn{3}{c}{25(11)}                          & \multicolumn{2}{c}{$\frac{1}{2}U_{11}$}    & \multicolumn{2}{c}{0}         & \multicolumn{2}{c}{0}         & 16(6)              &               \\
		& O(4)                             & 12k                              & \multicolumn{3}{c}{0.15478(17)}                                                                  & \multicolumn{2}{c}{0.3096(3)}                                                          & 0.05196(9)                                                & \multicolumn{2}{c}{1}                                       & 15(6)                  & \multicolumn{3}{c}{15(7)}                                    & \multicolumn{3}{c}{27(8)}                           & \multicolumn{2}{c}{$\frac{1}{2}U_{22}$}    & \multicolumn{2}{c}{$\frac{1}{2}U_{23}$}    & \multicolumn{2}{c}{3(5)}      & 19(5)              &               \\
		& O(5)                             & 4f                               & \multicolumn{3}{c}{$\frac{1}{3}$}                                                                            & \multicolumn{2}{c}{$\frac{2}{3}$}                                                                  & 0.55440(16)                                               & \multicolumn{2}{c}{1}                                       & 24(8)                  & \multicolumn{3}{c}{$U_{11}$}                                      & \multicolumn{3}{c}{24(13)}                          & \multicolumn{2}{c}{$\frac{1}{2}U_{11}$}    & \multicolumn{2}{c}{0}         & \multicolumn{2}{c}{0}         & 24(7)              &              
	\end{tabular}
\end{table}

\begin{table}[!h]
\renewcommand{\tablename}{Table S}
\caption{Details of the single crystal structure solution of EuAl$_{12}$O$_{19}$ using the I19 single crystal diffractometer at the Diamond Light Source, at $T=80$\,K}
\label{table:SCXRDtable80K}
        \begin{tabular}{cccccccccccccccccccccccccc}
		\hline
		
		\multicolumn{5}{c}{\begin{tabular}[c]{@{}c@{}}Space   group: \\    $P6_3/mmc$ (\#194, setting 1)\end{tabular}}                                                 & \multicolumn{5}{c}{$a$ = 5.56480(2) $\mathring{\mathrm{A}}$}                                                         & \multicolumn{5}{c}{$c$ = 21.9788(2)  $\mathring{\mathrm{A}}$}                                                                                                          & \multicolumn{6}{c}{$V$ = 589.4 $\mathring{\mathrm{A}}^3$}                                                                                                               & \multicolumn{5}{c}{$Z$ = 2}                                                                                                                           \\
		
		\hline
		
		\multicolumn{4}{c}{\begin{tabular}[c]{@{}c@{}}Radiation:   \\    X-rays synchrotron \\    ($\lambda$ = 0.6889 $\mathring{\mathrm{A}}$)\end{tabular}} & \multicolumn{3}{c}{\begin{tabular}[c]{@{}c@{}}Reflections   \\ collected/unique/used: \\ 13284/722/590\\\end{tabular}} & \multicolumn{3}{c}{\begin{tabular}[c]{@{}c@{}}Final $R$ indices: \\    $R$ = 3.22 \% \\    $wR_2$ = 8.84 \% \\  Parameters: 42\end{tabular}} & \multicolumn{7}{c}{\begin{tabular}[c]{@{}c@{}}Maximum   difference:\\  Peaks \& Holes \\        0.44 $e\mathring{\mathrm{A}}^{-3}$   \&   $-$0.62 $e\mathring{\mathrm{A}}^{-3}$ \end{tabular}} & \multicolumn{4}{c}{\begin{tabular}[c]{@{}c@{}}Density: \\    $\rho$ = 4.3933 g.cm$^{-3}$\end{tabular}} & \multicolumn{4}{c}{\begin{tabular}[c]{@{}c@{}}  Absorption \\ coefficient: \\    $\mu$ = 5.829 mm$^{-1}$\end{tabular}} \\
		
		\hline
		
		\multirow{2}{*}{$T$ (K)}             & \multirow{2}{*}{Atom}            & \multirow{2}{*}{Site}            & \multicolumn{3}{c}{\multirow{2}{*}{$x$}}                                                           & \multicolumn{2}{c}{\multirow{2}{*}{$y$}}                                                 & \multirow{2}{*}{$z$}                                        & \multicolumn{2}{c}{\multirow{2}{*}{Occ.}}                   & \multicolumn{15}{c}{$U_{\mathrm{ij}}$ ($\times$10$^4  \mathring{\mathrm{A}}^2$)}                                                                                                                                                                                                                                             \\ \cline{12-25}
		&                                  &                                  & \multicolumn{3}{c}{}                                                                             & \multicolumn{2}{c}{}                                                                   &                                                           & \multicolumn{2}{c}{}                                        & \multicolumn{2}{c}{$U_{11}$}                    & \multicolumn{2}{c}{$U_{22}$}                 & \multicolumn{2}{c}{$U_{33}$}              & \multicolumn{3}{c}{$U_{12}$}                          & \multicolumn{2}{c}{$U_{13}$}       & \multicolumn{2}{c}{$U_{23}$}       & \multicolumn{1}{c}{$U_{\mathrm{eq}}$}            &               \\
		\hline
		
		\multirow{11}{*}{80 K}             & Eu                               & 2d                               & \multicolumn{3}{c}{$\frac{1}{3}$}                                                                          & \multicolumn{2}{c}{$\frac{2}{3}$}                                                                  & $\frac{3}{4}$                                                      & \multicolumn{2}{c}{1}                                       & 30(1)                  & \multicolumn{3}{c}{$U_{11}$}                                      & \multicolumn{3}{c}{29(2)}                           & \multicolumn{2}{c}{$\frac{1}{2}U_{11}$}    & \multicolumn{2}{c}{0}         & \multicolumn{2}{c}{0}         & 30(1)              &               \\
		& Al(1)                            & 12k                              & \multicolumn{3}{c}{0.16833(4)}                                                                   & \multicolumn{2}{c}{0.33666(4)}                                                         & 0.60829(5)                                              & \multicolumn{2}{c}{1}                                       & 14(3)                  & \multicolumn{3}{c}{$U_{11}$}                                      & \multicolumn{3}{c}{16(4)}                           & \multicolumn{2}{c}{10(3)}      & \multicolumn{2}{c}{$-U_{23}$}      & \multicolumn{2}{c}{1(1)}      & 14(3)              &               \\
		& Al(2)                            & 4f                               & \multicolumn{3}{c}{$\frac{1}{3}$}                                                                            & \multicolumn{2}{c}{$\frac{2}{3}$}                                                                  & 0.19047(7)                                               & \multicolumn{2}{c}{1}                                       & 23(4)                  & \multicolumn{3}{c}{$U_{11}$}                                      & \multicolumn{3}{c}{10(6)}                           & \multicolumn{2}{c}{$\frac{1}{2}U_{11}$}    & \multicolumn{2}{c}{0}         & \multicolumn{2}{c}{0}         & 19(3)              &               \\
		& Al(3)                            & 2a                               & \multicolumn{3}{c}{0}                                                                            & \multicolumn{2}{c}{0}                                                                  & 0                                                         & \multicolumn{2}{c}{1}                                       & 20(4)                  & \multicolumn{3}{c}{$U_{11}$}                                      & \multicolumn{3}{c}{2(7)}                            & \multicolumn{2}{c}{$\frac{1}{2}U_{11}$}    & \multicolumn{2}{c}{0}         & \multicolumn{2}{c}{0}         & 14(4)              &               \\
		& Al(4)                            & 4f                               & \multicolumn{3}{c}{$\frac{1}{3}$}                                                                            & \multicolumn{2}{c}{$\frac{2}{3}$}                                                                  & 0.02838(7)                                           & \multicolumn{2}{c}{1}                                       & 12(4)                  & \multicolumn{3}{c}{$U_{11}$}                                      & \multicolumn{3}{c}{30(8)}                           & \multicolumn{2}{c}{$\frac{1}{2}U_{11}$}    & \multicolumn{2}{c}{0}         & \multicolumn{2}{c}{0}         & 18(4)              &               \\
		& Al(5)                            & 4e                               & \multicolumn{3}{c}{0}                                                                            & \multicolumn{2}{c}{0}                                                                  & 0.24043(12)                                              & \multicolumn{2}{c}{$\frac{1}{2}$}                                       & 17(5)                  & \multicolumn{3}{c}{$U_{11}$}                                      & \multicolumn{3}{c}{2(20)}                           & \multicolumn{2}{c}{$\frac{1}{2}U_{11}$}    & \multicolumn{2}{c}{0}         & \multicolumn{2}{c}{0}         & 12(7)              &               \\
		& O(1)                             & 6h                               & \multicolumn{3}{c}{0.18115(18)}                                                                    & \multicolumn{2}{c}{0.3623(4)}                                                          & $\frac{1}{4}$                                                         & \multicolumn{2}{c}{1}                                       & 28(7)                  & \multicolumn{3}{c}{14(6)}                                   & \multicolumn{3}{c}{45(9)}                          & \multicolumn{2}{c}{$\frac{1}{2}U_{22}$}    & \multicolumn{2}{c}{0}         & \multicolumn{2}{c}{0}         & 31(5)              &               \\
		& O(2)                             & 12k                              & \multicolumn{3}{c}{0.50232(8)}                                                                    & \multicolumn{2}{c}{0.00464(16)}                                                        & 0.14809(8)                                                & \multicolumn{2}{c}{1}                                       & 28(6)                  & \multicolumn{3}{c}{25(5)}                                    & \multicolumn{3}{c}{17(6)}                           & \multicolumn{2}{c}{$\frac{1}{2}U_{11}$}    & \multicolumn{2}{c}{2$U_{23}$}      & \multicolumn{2}{c}{1(2)}      & 23(4)              &               \\
		& O(3)                             & 4e                               & \multicolumn{3}{c}{0}                                                                            & \multicolumn{2}{c}{0}                                                                  & 0.14846(11)                                              & \multicolumn{2}{c}{1}                                       & 28(6)                  & \multicolumn{3}{c}{$U_{11}$}                                      & \multicolumn{3}{c}{43(9)}                          & \multicolumn{2}{c}{$\frac{1}{2}U_{11}$}    & \multicolumn{2}{c}{0}         & \multicolumn{2}{c}{0}         & 33(5)              &               \\
		& O(4)                             & 12k                              & \multicolumn{3}{c}{0.15490(12)}                                                                  & \multicolumn{2}{c}{0.3098(2)}                                                          & 0.05170(10)                                                & \multicolumn{2}{c}{1}                                       & 24(5)                  & \multicolumn{3}{c}{15(5)}                                    & \multicolumn{3}{c}{31(7)}                           & \multicolumn{2}{c}{$\frac{1}{2}U_{22}$}    & \multicolumn{2}{c}{$\frac{1}{2}U_{23}$}    & \multicolumn{2}{c}{5(4)}      & 25(4)              &               \\
		& O(5)                             & 4f                               & \multicolumn{3}{c}{$\frac{1}{3}$}                                                                            & \multicolumn{2}{c}{$\frac{2}{3}$}                                                                  & 0.55428(16)                                              & \multicolumn{2}{c}{1}                                       & 32(8)                & \multicolumn{3}{c}{$U_{11}$}                                      & \multicolumn{3}{c}{16(13)}                          & \multicolumn{2}{c}{$\frac{1}{2}U_{11}$}    & \multicolumn{2}{c}{0}         & \multicolumn{2}{c}{0}         & 27(7)              &              
	\end{tabular}
\end{table}

\begin{table}[!h]
\renewcommand{\tablename}{Table S}
\caption{Details of the single crystal structure solution of EuAl$_{12}$O$_{19}$ using the I19 single crystal diffractometer at the Diamond Light Source, at $T=120$\,K}
\label{table:SCXRDtable120K}
        \begin{tabular}{cccccccccccccccccccccccccc}
		\hline
		
		\multicolumn{5}{c}{\begin{tabular}[c]{@{}c@{}}Space   group: \\    $P6_3/mmc$ (\#194, setting 1)\end{tabular}}                                                 & \multicolumn{5}{c}{$a$ = 5.56620(18)  $\mathring{\mathrm{A}}$}                                                         & \multicolumn{5}{c}{$c$ = 21.9871(1)   $\mathring{\mathrm{A}}$}                                                                                                          & \multicolumn{6}{c}{$V$ = 590.0 $\mathring{\mathrm{A}}^3$}                                                                                                               & \multicolumn{5}{c}{$Z$ = 2}                                                                                                                           \\
		
		\hline
		
		\multicolumn{4}{c}{\begin{tabular}[c]{@{}c@{}}Radiation:   \\    X-rays synchrotron \\    ($\lambda$ = 0.6889 $\mathring{\mathrm{A}}$)\end{tabular}} & \multicolumn{3}{c}{\begin{tabular}[c]{@{}c@{}}Reflections   \\ collected/unique/used: \\ 13554/725/613\\\end{tabular}} & \multicolumn{3}{c}{\begin{tabular}[c]{@{}c@{}}Final $R$ indices: \\    $R$ = 1.77 \% \\    $wR_2$ = 5.71 \% \\  Parameters: 42\end{tabular}} & \multicolumn{7}{c}{\begin{tabular}[c]{@{}c@{}}Maximum   difference:\\  Peaks \& Holes \\        0.25 $e\mathring{\mathrm{A}}^{-3}$   \&   $-$0.51 $e\mathring{\mathrm{A}}^{-3}$ \end{tabular}} & \multicolumn{4}{c}{\begin{tabular}[c]{@{}c@{}}Density: \\    $\rho$ = 4.3894 g.cm$^{-3}$\end{tabular}} & \multicolumn{4}{c}{\begin{tabular}[c]{@{}c@{}}  Absorption \\ coefficient: \\    $\mu$ = 5.824 mm$^{-1}$\end{tabular}} \\
		
		\hline
		
		\multirow{2}{*}{$T$ (K)}             & \multirow{2}{*}{Atom}            & \multirow{2}{*}{Site}            & \multicolumn{3}{c}{\multirow{2}{*}{$x$}}                                                           & \multicolumn{2}{c}{\multirow{2}{*}{$y$}}                                                 & \multirow{2}{*}{$z$}                                        & \multicolumn{2}{c}{\multirow{2}{*}{Occ.}}                   & \multicolumn{15}{c}{$U_{\mathrm{ij}}$ ($\times$10$^4  \mathring{\mathrm{A}}^2$)}                                                                                                                                                                                                                                             \\ \cline{12-25}
		&                                  &                                  & \multicolumn{3}{c}{}                                                                             & \multicolumn{2}{c}{}                                                                   &                                                           & \multicolumn{2}{c}{}                                        & \multicolumn{2}{c}{$U_{11}$}                    & \multicolumn{2}{c}{$U_{22}$}                 & \multicolumn{2}{c}{$U_{33}$}              & \multicolumn{3}{c}{$U_{12}$}                          & \multicolumn{2}{c}{$U_{13}$}       & \multicolumn{2}{c}{$U_{23}$}       & \multicolumn{1}{c}{$U_{\mathrm{eq}}$}            &               \\
		\hline
		
		\multirow{11}{*}{120 K}             & Eu                               & 2d                               & \multicolumn{3}{c}{$\frac{1}{3}$}                                                                          & \multicolumn{2}{c}{$\frac{2}{3}$}                                                                  & $\frac{3}{4}$                                                      & \multicolumn{2}{c}{1}                                       & 37(1)                  & \multicolumn{3}{c}{$U_{11}$}                                      & \multicolumn{3}{c}{34(1)}                           & \multicolumn{2}{c}{$\frac{1}{2}U_{11}$}    & \multicolumn{2}{c}{0}         & \multicolumn{2}{c}{0}         & 36(1)              &               \\
		& Al(1)                            & 12k                              & \multicolumn{3}{c}{0.16830(4)}                                                                   & \multicolumn{2}{c}{0.33660(4)}                                                         & 0.60834(3)                                             & \multicolumn{2}{c}{1}                                       & 21(2)                  & \multicolumn{3}{c}{$U_{11}$}                                      & \multicolumn{3}{c}{20(3)}                           & \multicolumn{2}{c}{11(2)}      & \multicolumn{2}{c}{$-U_{23}$}      & \multicolumn{2}{c}{1(1)}      & 21(2)              &               \\
		& Al(2)                            & 4f                               & \multicolumn{3}{c}{$\frac{1}{3}$}                                                                            & \multicolumn{2}{c}{$\frac{2}{3}$}                                                                  & 0.19058(5)                                              & \multicolumn{2}{c}{1}                                       & 25(2)                  & \multicolumn{3}{c}{$U_{11}$}                                      & \multicolumn{3}{c}{8(4)}                           & \multicolumn{2}{c}{$\frac{1}{2}U_{11}$}    & \multicolumn{2}{c}{0}         & \multicolumn{2}{c}{0}         & 20(2)              &               \\
		& Al(3)                            & 2a                               & \multicolumn{3}{c}{0}                                                                            & \multicolumn{2}{c}{0}                                                                  & 0                                                         & \multicolumn{2}{c}{1}                                       & 21(3)                  & \multicolumn{3}{c}{$U_{11}$}                                      & \multicolumn{3}{c}{14(5)}                            & \multicolumn{2}{c}{$\frac{1}{2}U_{11}$}    & \multicolumn{2}{c}{0}         & \multicolumn{2}{c}{0}         & 19(2)              &               \\
		& Al(4)                            & 4f                               & \multicolumn{3}{c}{$\frac{1}{3}$}                                                                            & \multicolumn{2}{c}{$\frac{2}{3}$}                                                                  & 0.02832(5)                                           & \multicolumn{2}{c}{1}                                       & 24(3)                  & \multicolumn{3}{c}{$U_{11}$}                                      & \multicolumn{3}{c}{21(4)}                           & \multicolumn{2}{c}{$\frac{1}{2}U_{11}$}    & \multicolumn{2}{c}{0}         & \multicolumn{2}{c}{0}         & 23(2)              &               \\
		& Al(5)                            & 4e                               & \multicolumn{3}{c}{0}                                                                            & \multicolumn{2}{c}{0}                                                                  & 0.24054(7)                                            & \multicolumn{2}{c}{$\frac{1}{2}$}                                       & 21(3)                  & \multicolumn{3}{c}{$U_{11}$}                                      & \multicolumn{3}{c}{12(14)}                           & \multicolumn{2}{c}{$\frac{1}{2}U_{11}$}    & \multicolumn{2}{c}{0}         & \multicolumn{2}{c}{0}         & 18(5)              &               \\
		& O(1)                             & 6h                               & \multicolumn{3}{c}{0.18132(15)}                                                                    & \multicolumn{2}{c}{0.36260(3)}                                                          & $\frac{1}{4}$                                                         & \multicolumn{2}{c}{1}                                       & 47(6)                  & \multicolumn{3}{c}{13(6)}                                   & \multicolumn{3}{c}{43(7)}                          & \multicolumn{2}{c}{$\frac{1}{2}U_{22}$}    & \multicolumn{2}{c}{0}         & \multicolumn{2}{c}{0}         & 38(4)              &               \\
		& O(2)                             & 12k                              & \multicolumn{3}{c}{0.50250(8)}                                                                    & \multicolumn{2}{c}{0.00499(17)}                                                        & 0.14809(5)                                               & \multicolumn{2}{c}{1}                                       & 30(4)                  & \multicolumn{3}{c}{25(3)}                                    & \multicolumn{3}{c}{25(4)}                           & \multicolumn{2}{c}{$\frac{1}{2}U_{11}$}    & \multicolumn{2}{c}{2$U_{23}$}      & \multicolumn{2}{c}{1(1)}      & 26(3)              &               \\
		& O(3)                             & 4e                               & \multicolumn{3}{c}{0}                                                                            & \multicolumn{2}{c}{0}                                                                  & 0.14841(8)                                              & \multicolumn{2}{c}{1}                                       & 25(4)                  & \multicolumn{3}{c}{$U_{11}$}                                      & \multicolumn{3}{c}{22(6)}                          & \multicolumn{2}{c}{$\frac{1}{2}U_{11}$}    & \multicolumn{2}{c}{0}         & \multicolumn{2}{c}{0}         & 24(4)              &               \\
		& O(4)                             & 12k                              & \multicolumn{3}{c}{0.15491(10)}                                                                  & \multicolumn{2}{c}{0.30980(2)}                                                          & 0.05185(6)                                               & \multicolumn{2}{c}{1}                                       & 28(4)                  & \multicolumn{3}{c}{25(4)}                                    & \multicolumn{3}{c}{22(5)}                           & \multicolumn{2}{c}{$\frac{1}{2}U_{22}$}    & \multicolumn{2}{c}{$\frac{1}{2}U_{23}$}    & \multicolumn{2}{c}{2(3)}      & 25(3)              &               \\
		& O(5)                             & 4f                               & \multicolumn{3}{c}{$\frac{1}{3}$}                                                                            & \multicolumn{2}{c}{$\frac{2}{3}$}                                                                  & 0.55422(10)                                             & \multicolumn{2}{c}{1}                                       & 30(5)                & \multicolumn{3}{c}{$U_{11}$}                                      & \multicolumn{3}{c}{3(8)}                          & \multicolumn{2}{c}{$\frac{1}{2}U_{11}$}    & \multicolumn{2}{c}{0}         & \multicolumn{2}{c}{0}         & 21(4)              &              
	\end{tabular}
\end{table}

\begin{table}[!h]
\renewcommand{\tablename}{Table S}
\caption{Details of the single crystal structure solution of EuAl$_{12}$O$_{19}$ using the I19 single crystal diffractometer at the Diamond Light Source, at $T=225$\,K}
\label{table:SCXRDtable225K}
        \begin{tabular}{cccccccccccccccccccccccccc}
		\hline
		
				\multicolumn{5}{c}{\begin{tabular}[c]{@{}c@{}}Space   group: \\    $P6_3/mmc$ (\#194, setting 1)\end{tabular}}                                                 & \multicolumn{5}{c}{$a$ = 5.56800(2)   $\mathring{\mathrm{A}}$}                                                         & \multicolumn{5}{c}{$c$ = 21.9997(1)    $\mathring{\mathrm{A}}$ }                                                                                                          & \multicolumn{6}{c}{$V$ = 590.7 $\mathring{\mathrm{A}}^3$}                                                                                                               & \multicolumn{5}{c}{$Z$ = 2}                                                                                                                           \\
		
		\hline
		
		\multicolumn{4}{c}{\begin{tabular}[c]{@{}c@{}}Radiation:   \\    X-rays synchrotron \\    ($\lambda$ = 0.6889 $\mathring{\mathrm{A}}$)\end{tabular}} & \multicolumn{3}{c}{\begin{tabular}[c]{@{}c@{}}Reflections   \\ collected/unique/used: \\ 13176/725/621\\\end{tabular}} & \multicolumn{3}{c}{\begin{tabular}[c]{@{}c@{}}Final $R$ indices: \\    $R$ = 1.85 \% \\    $wR_2$ = 5.58 \% \\  Parameters: 42\end{tabular}} & \multicolumn{7}{c}{\begin{tabular}[c]{@{}c@{}}Maximum   difference:\\  Peaks \& Holes \\        0.74 $e\mathring{\mathrm{A}}^{-3}$   \&   $-$1.24 $e\mathring{\mathrm{A}}^{-3}$ \end{tabular}} & \multicolumn{4}{c}{\begin{tabular}[c]{@{}c@{}}Density: \\    $\rho$ = 4.3840 g.cm$^{-3}$\end{tabular}} & \multicolumn{4}{c}{\begin{tabular}[c]{@{}c@{}}  Absorption \\ coefficient: \\    $\mu$ = 5.817 mm$^{-1}$\end{tabular}} \\
		
		\hline
		
		\multirow{2}{*}{$T$ (K)}             & \multirow{2}{*}{Atom}            & \multirow{2}{*}{Site}            & \multicolumn{3}{c}{\multirow{2}{*}{$x$}}                                                           & \multicolumn{2}{c}{\multirow{2}{*}{$y$}}                                                 & \multirow{2}{*}{$z$}                                        & \multicolumn{2}{c}{\multirow{2}{*}{Occ.}}                   & \multicolumn{15}{c}{$U_{\mathrm{ij}}$ ($\times$10$^4  \mathring{\mathrm{A}}^2$)}                                                                                                                                                                                                                                             \\ \cline{12-25}
		&                                  &                                  & \multicolumn{3}{c}{}                                                                             & \multicolumn{2}{c}{}                                                                   &                                                           & \multicolumn{2}{c}{}                                        & \multicolumn{2}{c}{$U_{11}$}                    & \multicolumn{2}{c}{$U_{22}$}                 & \multicolumn{2}{c}{$U_{33}$}              & \multicolumn{3}{c}{$U_{12}$}                          & \multicolumn{2}{c}{$U_{13}$}       & \multicolumn{2}{c}{$U_{23}$}       & \multicolumn{1}{c}{$U_{\mathrm{eq}}$}            &               \\
		\hline
		
		\multirow{11}{*}{225 K}             & Eu                               & 2d                               & \multicolumn{3}{c}{$\frac{1}{3}$}                                                                          & \multicolumn{2}{c}{$\frac{2}{3}$}                                                                  & $\frac{3}{4}$                                                      & \multicolumn{2}{c}{1}                                       & 51(1)                  & \multicolumn{3}{c}{$U_{11}$}                                      & \multicolumn{3}{c}{50(1)}                           & \multicolumn{2}{c}{$\frac{1}{2}U_{11}$}    & \multicolumn{2}{c}{0}         & \multicolumn{2}{c}{0}         & 50(1)              &               \\
		& Al(1)                            & 12k                              & \multicolumn{3}{c}{0.16836(4)}                                                                   & \multicolumn{2}{c}{0.33671(8)}                                                         & 0.60836(3)                                             & \multicolumn{2}{c}{1}                                       & 27(2)                  & \multicolumn{3}{c}{$U_{11}$}                                      & \multicolumn{3}{c}{27(3)}                           & \multicolumn{2}{c}{11(1)}      & \multicolumn{2}{c}{$-U_{23}$}      & \multicolumn{2}{c}{3(1)}      & 26(2)              &               \\
		& Al(2)                            & 4f                               & \multicolumn{3}{c}{$\frac{1}{3}$}                                                                            & \multicolumn{2}{c}{$\frac{2}{3}$}                                                                  & 0.19052(5)                                             & \multicolumn{2}{c}{1}                                       & 24(2)                  & \multicolumn{3}{c}{$U_{11}$}                                      & \multicolumn{3}{c}{21(4)}                           & \multicolumn{2}{c}{$\frac{1}{2}U_{11}$}    & \multicolumn{2}{c}{0}         & \multicolumn{2}{c}{0}         & 23(2)              &               \\
		& Al(3)                            & 2a                               & \multicolumn{3}{c}{0}                                                                            & \multicolumn{2}{c}{0}                                                                  & 0                                                         & \multicolumn{2}{c}{1}                                       & 28(3)                  & \multicolumn{3}{c}{$U_{11}$}                                      & \multicolumn{3}{c}{25(4)}                            & \multicolumn{2}{c}{$\frac{1}{2}U_{11}$}    & \multicolumn{2}{c}{0}         & \multicolumn{2}{c}{0}         & 25(2)              &               \\
		& Al(4)                            & 4f                               & \multicolumn{3}{c}{$\frac{1}{3}$}                                                                            & \multicolumn{2}{c}{$\frac{2}{3}$}                                                                  & 0.02829(4)                                           & \multicolumn{2}{c}{1}                                       & 25(3)                  & \multicolumn{3}{c}{$U_{11}$}                                      & \multicolumn{3}{c}{23(1)}                           & \multicolumn{2}{c}{$\frac{1}{2}U_{11}$}    & \multicolumn{2}{c}{0}         & \multicolumn{2}{c}{0}         & 25(2)              &               \\
		& Al(5)                            & 4e                               & \multicolumn{3}{c}{0}                                                                            & \multicolumn{2}{c}{0}                                                                  & 0.24057(7)                                            & \multicolumn{2}{c}{$\frac{1}{2}$}                                       & 26(3)                  & \multicolumn{3}{c}{$U_{11}$}                                      & \multicolumn{3}{c}{23(13)}                           & \multicolumn{2}{c}{$\frac{1}{2}U_{11}$}    & \multicolumn{2}{c}{0}         & \multicolumn{2}{c}{0}         & 25(5)              &               \\
		& O(1)                             & 6h                               & \multicolumn{3}{c}{0.1813(3)}                                                                    & \multicolumn{2}{c}{0.36260(15)}                                                          & $\frac{1}{4}$                                                         & \multicolumn{2}{c}{1}                                       & 21(6)                  & \multicolumn{3}{c}{56(6)}                                   & \multicolumn{3}{c}{42(7)}                          & \multicolumn{2}{c}{$\frac{1}{2}U_{22}$}    & \multicolumn{2}{c}{0}         & \multicolumn{2}{c}{0}         & 44(4)              &               \\
		& O(2)                             & 12k                              & \multicolumn{3}{c}{0.50242(9)}                                                                    & \multicolumn{2}{c}{0.00484(9)}                                                        & 0.14811(5)                                               & \multicolumn{2}{c}{1}                                       & 29(3)                  & \multicolumn{3}{c}{29(3)}                                    & \multicolumn{3}{c}{37(4)}                           & \multicolumn{2}{c}{$\frac{1}{2}U_{11}$}    & \multicolumn{2}{c}{2$U_{23}$}      & \multicolumn{2}{c}{5(1)}      & 33(3)             &               \\
		& O(3)                             & 4e                               & \multicolumn{3}{c}{0}                                                                            & \multicolumn{2}{c}{0}                                                                  & 0.14831(8)                                             & \multicolumn{2}{c}{1}                                       & 23(4)                  & \multicolumn{3}{c}{$U_{11}$}                                      & \multicolumn{3}{c}{44(7)}                          & \multicolumn{2}{c}{$\frac{1}{2}U_{11}$}    & \multicolumn{2}{c}{0}         & \multicolumn{2}{c}{0}         & 30(4)             &               \\
		& O(4)                             & 12k                              & \multicolumn{3}{c}{0.1551(2)}                                                                  & \multicolumn{2}{c}{0.31020(10)}                                                          & 0.05176(6)                                              & \multicolumn{2}{c}{1}                                       & 16(4)            & \multicolumn{3}{c}{25(4)}                                    & \multicolumn{3}{c}{32(5)}                           & \multicolumn{2}{c}{$\frac{1}{2}U_{22}$}    & \multicolumn{2}{c}{$\frac{1}{2}U_{23}$}    & \multicolumn{2}{c}{2(3)}      & 27(3)              &               \\
		& O(5)                             & 4f                               & \multicolumn{3}{c}{$\frac{1}{3}$}                                                                            & \multicolumn{2}{c}{$\frac{2}{3}$}                                                                  & 0.55435(10)                                             & \multicolumn{2}{c}{1}                                       & 31(5)                & \multicolumn{3}{c}{$U_{11}$}                                      & \multicolumn{3}{c}{17(8)}                          & \multicolumn{2}{c}{$\frac{1}{2}U_{11}$}    & \multicolumn{2}{c}{0}         & \multicolumn{2}{c}{0}         & 26(4)              &              
	\end{tabular}
\end{table}

\newpage

\section{Electron distribution map extracted by the Maximum of Entropy Method (MEM)}

\begin{figure*}[t]
\renewcommand{\figurename}{Figure S}
\begin{center}
\includegraphics[width=1\linewidth]{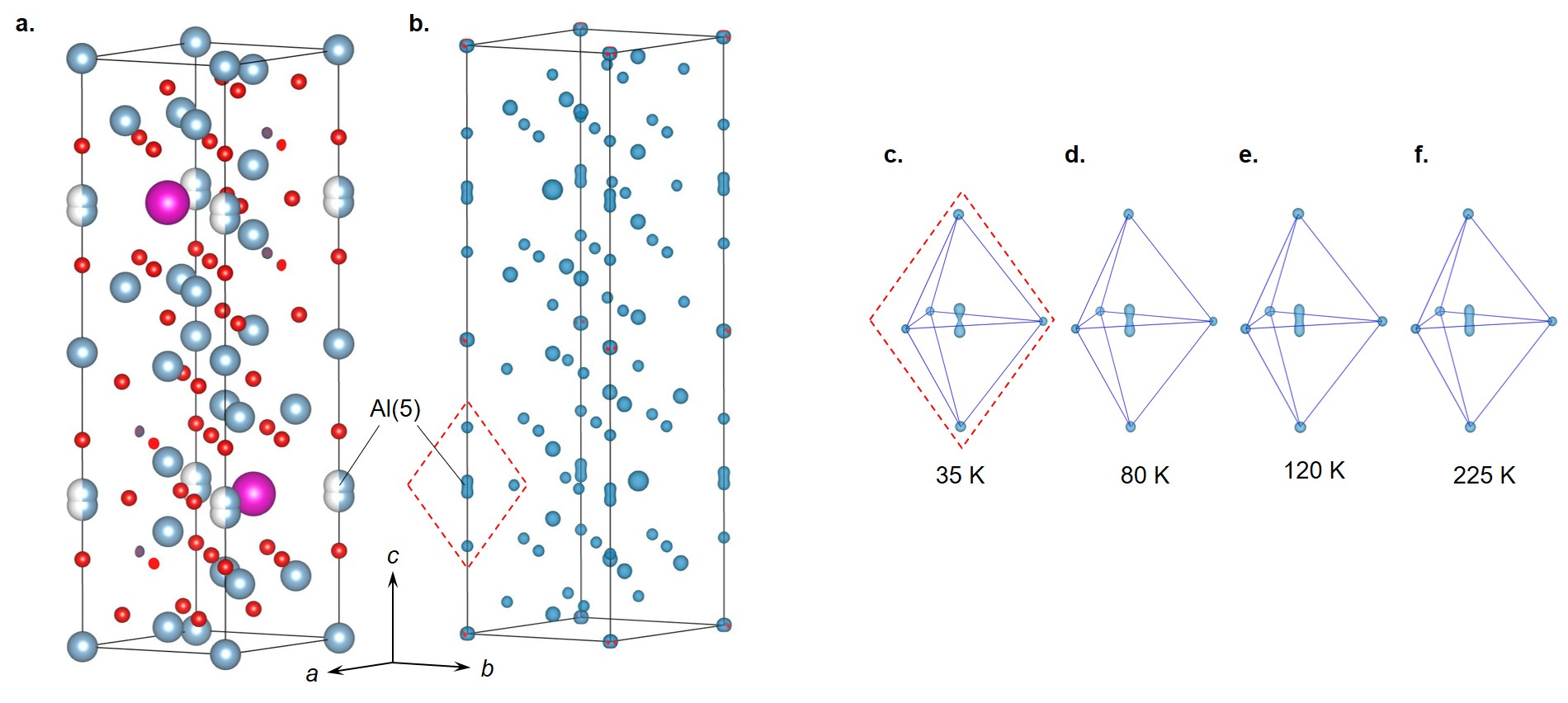}
\end{center}
\caption{Side-by-side comparison of the EuAl$_{12}$O$_{19}$ single crystal structure solution (a) at $T=35\,$K, with an isosurface plot of the MEM-calculated electron density distribution map (b) at an isosurface value of 3.5\,$e \mathring{\mathrm{A}}^{-3}$. The off-mirror-plane displacement of the Al(5) site is observed in the MEM-calculation in agreement with the split-site model used for the refinement. (c-f) The temperature-dependent changes of the Al(5) electron density distribution, as well as those of the surrounding bipyramid coordinating oxygen sites, with an isosurface level of 8.9\,$e \mathring{\mathrm{A}}^{-3}$. The apparent electron density bridging the two sites reduces on cooling from 225 K to 35 K, with no evident distortion of the oxygen atoms that create the coordination cage.}
\label{MEMsup}
\end{figure*}

\begin{figure*}[t]
\renewcommand{\figurename}{Figure S}
\begin{center}
\includegraphics[width=1\linewidth]{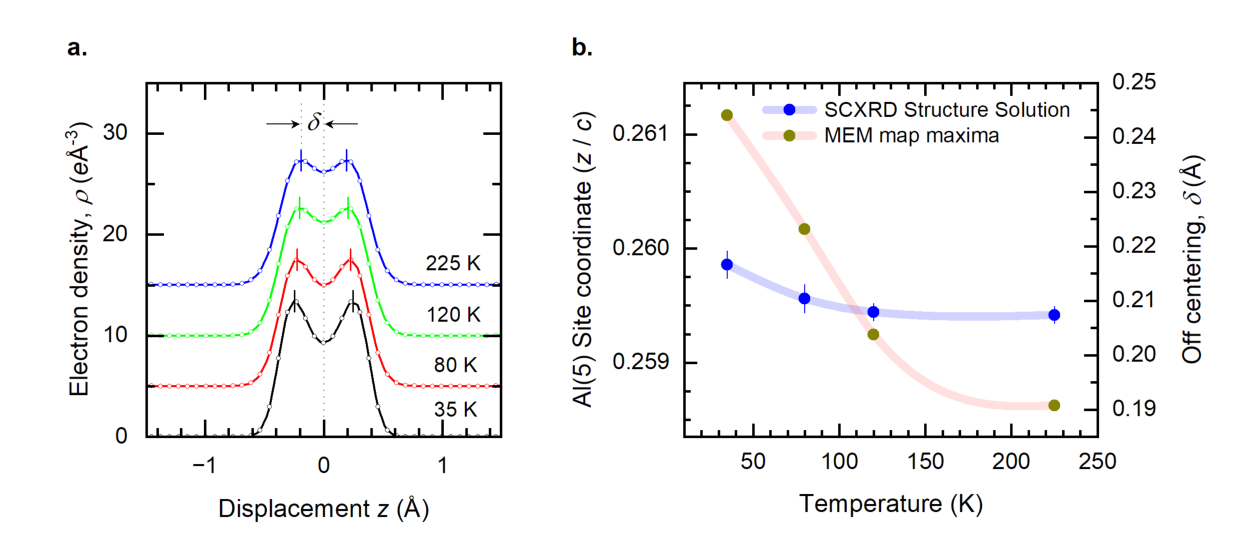}
\end{center}
\caption{a. Electron density as a one-dimensional cut along the $c$ axis bisecting the Al(5) site ($x$=0, $y$=0, z). The curves are offset for clarity. The off-centering $\delta$ of the Al(5) ion with respect to its oxygen cage is defined as indicated in the figure as the distance between the maximum of the electron density and the center of the bipyramid. (b.) Comparison of the temperature dependence of the refined atomic $z$-coordinate of the Al(5) ion with the maximum in the electron density map of the MEM distribution. The right axis shows absolute values of the off-centering distance $\delta$.}
\label{Cut}
\end{figure*}

The electron distribution map was obtained by the Maximum of Entropy Method (MEM) using the Dysnomia software suite~\cite{Momma2013} and it is represented in Fig.~S\ref{MEMsup} together with the crystal structure. Anomalous atomic displacement is observed only for the Al(5) ions, thus it is reasonable to assume that the measured dielectric relaxations in EuAl$_{12}$O$_{19}$ are mainly related to this particular site.

Cuts of the electron density along the diagonal of the bipyramid AlO$_5$ at different temperatures are shown in Fig.~S\ref{Cut}a. They unambiguously show the splitting of the Al(5) site into two off-centered positions at all temperatures. The temperature dependence of the off-centering of the Al(5) ion $\delta$ from structure solution and from MEM cuts are represented in Fig.~S\ref{Cut}b. Both methods show an enhancement of the off-centering upon cooling. This evolution can be explained from the reduction of the occupancy of the center of the bipyramid upon cooling 
and it does not imply any evolution of the double well potential itself with temperature. Thus the double well potential can be considered as stable in temperature and the distance between the center of the bipyramid and the ground state position of the Al(5) is best estimated from the 
lowest temperature MEM analysis ($T=35\,$K) at $\delta \approx 0.24\,$\AA.  The difference between Al(5) position determined from structure solution and the maxima position within the MEM maps results from the assumptions built into single crystal solution software assuming a symmetric thermal displacement ellipsoid, which must be skewed towards the mirror plane due to additional electron density between the sites.

\section{Temperature dependence of lattice parameters}

\begin{figure*}[h]
\renewcommand{\figurename}{Figure S}
\begin{minipage}{0.49\linewidth}
\begin{center}
\includegraphics[width=1\linewidth]{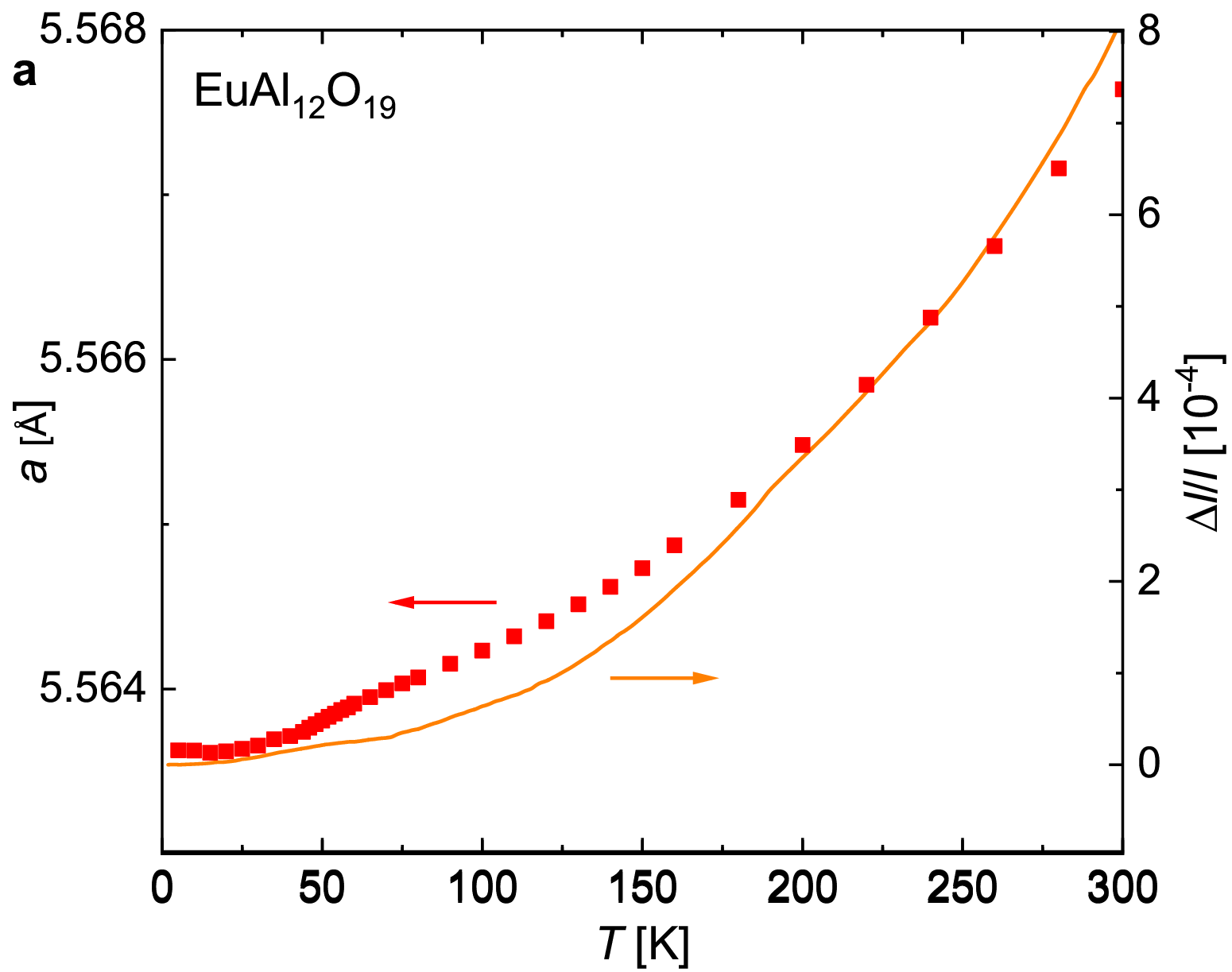}
\end{center}
\end{minipage}
\hfill
\begin{minipage}{0.49\linewidth}
\begin{center}
\includegraphics[width=1\linewidth]{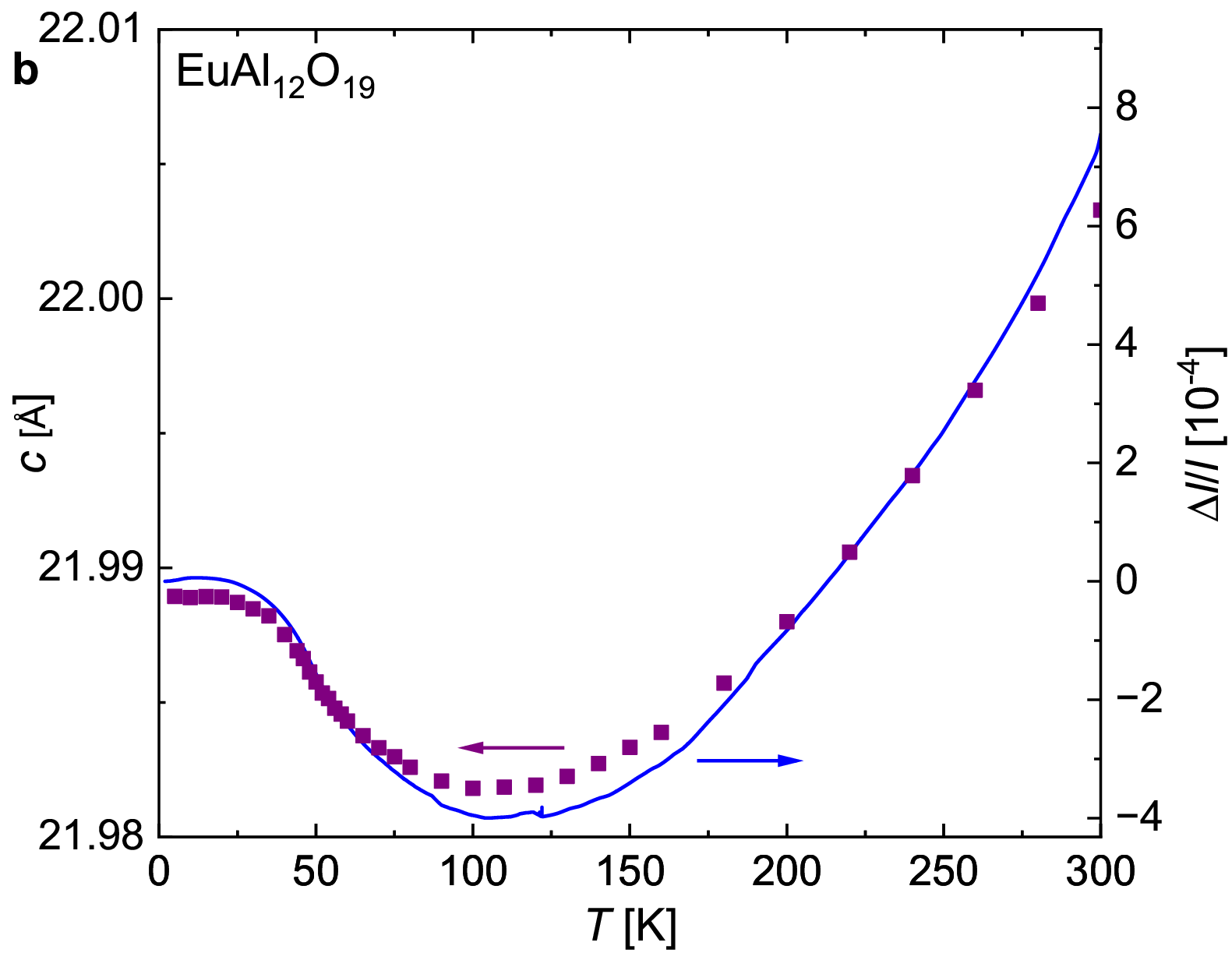}
\end{center}
\end{minipage}
\begin{center}
\includegraphics[width=0.5\linewidth]{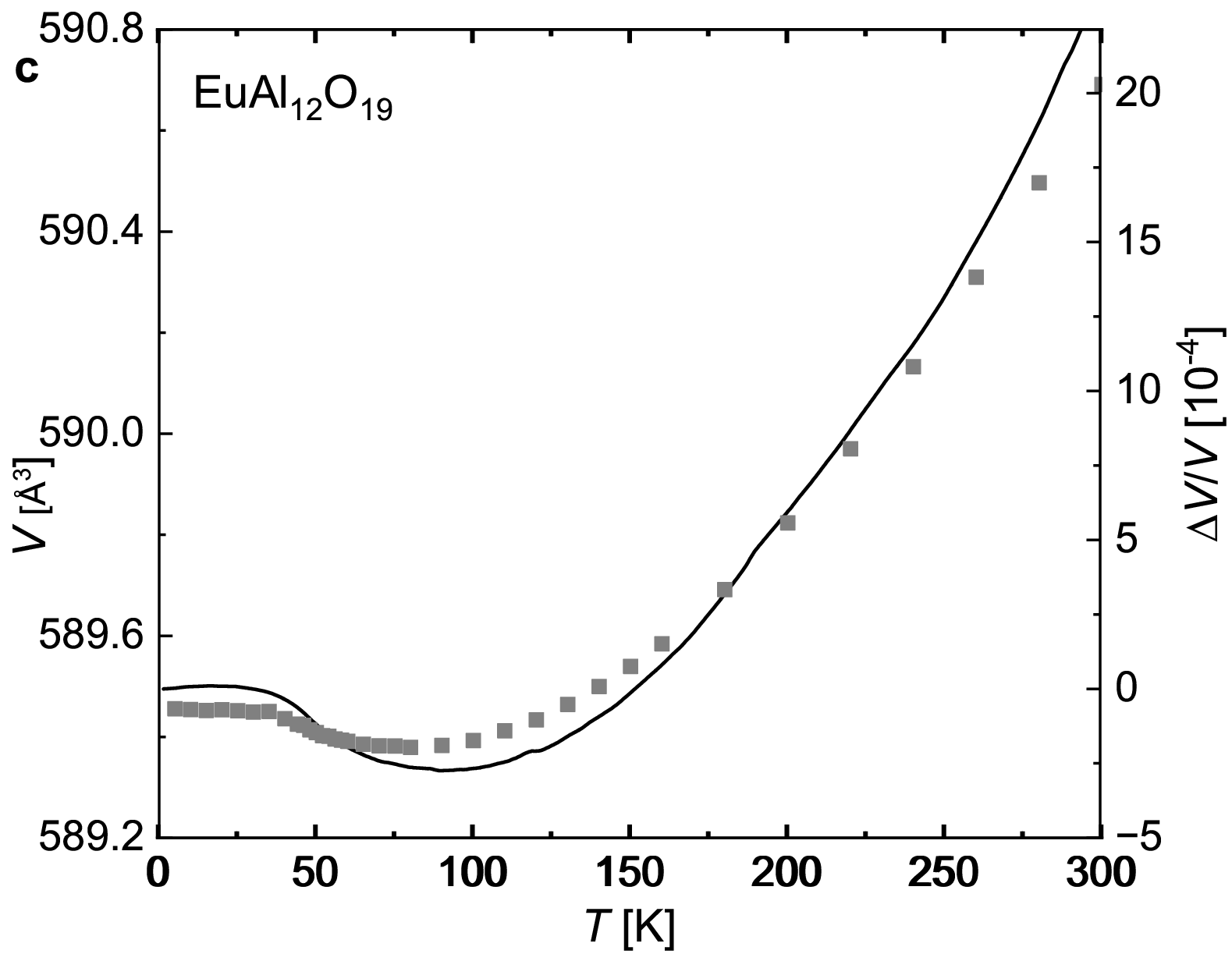}
\end{center}
\caption{Temperature dependence of the lattice parameter $a$ in \textbf{a}, the lattice parameter $c$ in \textbf{b} and the cell volume in \textbf{c} measured by single crystal x-ray diffraction. The relative change of lattice parameters and cell volume extracted from thermal expansion measurements are plotted together to show the agreement between both methods.}
\label{avsT}
\end{figure*}

The temperature dependence of lattice parameters was determined from single crystal diffraction by following the shift of selected reflections (Fig.~S\ref{avsT}). These measurements were performed down to $T=5\,$K using a laboratory x-ray source (see Methods section). The variation of lattice parameters and cell volume are in good agreement with thermal expansion results, especially both methods agree on the occurrence of negative thermal expansion along the $c$ axis between $T \approx 10\,$K and $T \approx 100\,$K. This negative thermal expansion along the $c$ axis exceeds the positive thermal expansion in the $ab$ plane implying an expansion of the crystal volume upon cooling (Fig.~S\ref{avsT}c).

\section{Complex dielectric permittivity in the THz range in the $\textbf{E}^{\omega}$$\parallel$~$\textbf{a}$ polarization.}

\begin{figure}[h!]
\renewcommand{\figurename}{Figure S}
	\begin{center}
		\includegraphics[width=1\linewidth]{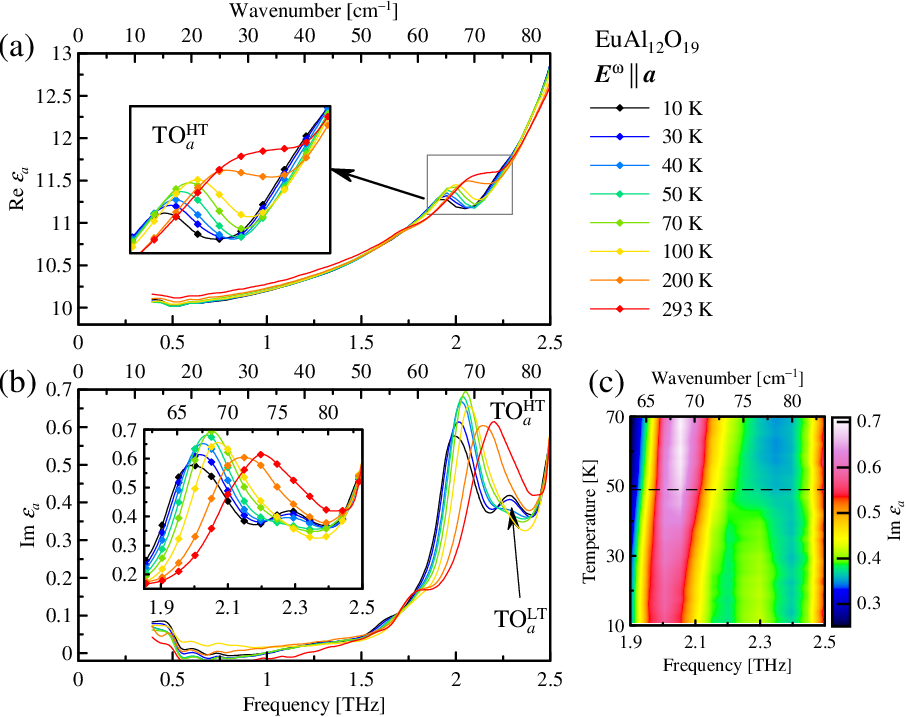}
		\caption{Temperature evolution of the (a) real and (b) imaginary part of THz complex permittivity spectra of EuAl$_{12}$O$_{19}$ measured in the $\textbf{E}^{\omega}$$\parallel$~$\textbf{a}$ configuration. Insets show details on TO$_a^\mathrm{HT}$  and TO$_a^\mathrm{LT}$ modes. (c) Colormap representation of the evolution of imaginary permittivity spectra between 10 and 70\,K.}
		\label{fig:THz_eps_Eparala}
	\end{center}
\end{figure}

The spectra of THz complex permittivity directly calculated from the measured complex transmittance spectra obtained in the $\textbf{E}^{\omega}$$\parallel$~$\textbf{a}$ polarization complete the image of phonon dynamics in a range from 13\,cm$^{-1}$ (i.e. 0.4\,THz) to 80\,cm$^{-1}$ (i.e. 2.5\,THz) (Fig.~S\ref{fig:THz_eps_Eparala}). It does not show any sign of the dielectric relaxation R1 observed in the $\textbf{E}^{\omega}$$\parallel$~$\textbf{c}$ THz spectra (Fig.~4c-d in the main text) confirming the unidirectional nature of this relaxation ascribed to the Ising-like electric dipoles AlO$_5$. A new mode TO$_a^\mathrm{LT}$ emerges on the verge of the THz range near 76\,cm$^{-1}$ below $T_S$ and thus supports the change of crystal symmetry on the local scale. Moreover, strongly anharmonic mode TO$_a^\mathrm{HT}$ dominates the spectra 
in the whole temperature range: it gradually softens from about 73\,cm$^{-1}$ at room temperature to 66\,cm$^{-1}$ at 10\,K without any critical behavior near the phase transition at $T_S=49$\,K. In a way, it may act as a counterpart to the TO$_c^\mathrm{HT}$  phonons, which harden substantially in the $\textbf{E}^{\omega}$$\parallel$~$\textbf{c}$ polarized IR spectra (See Fig.~4b in the main text).

\section{Confirmation of anisotropic hardening upon cooling from sound velocity experiments}

\begin{figure}[h!]
\renewcommand{\figurename}{Figure S}
	\begin{center}
		\includegraphics[width=0.6\linewidth]{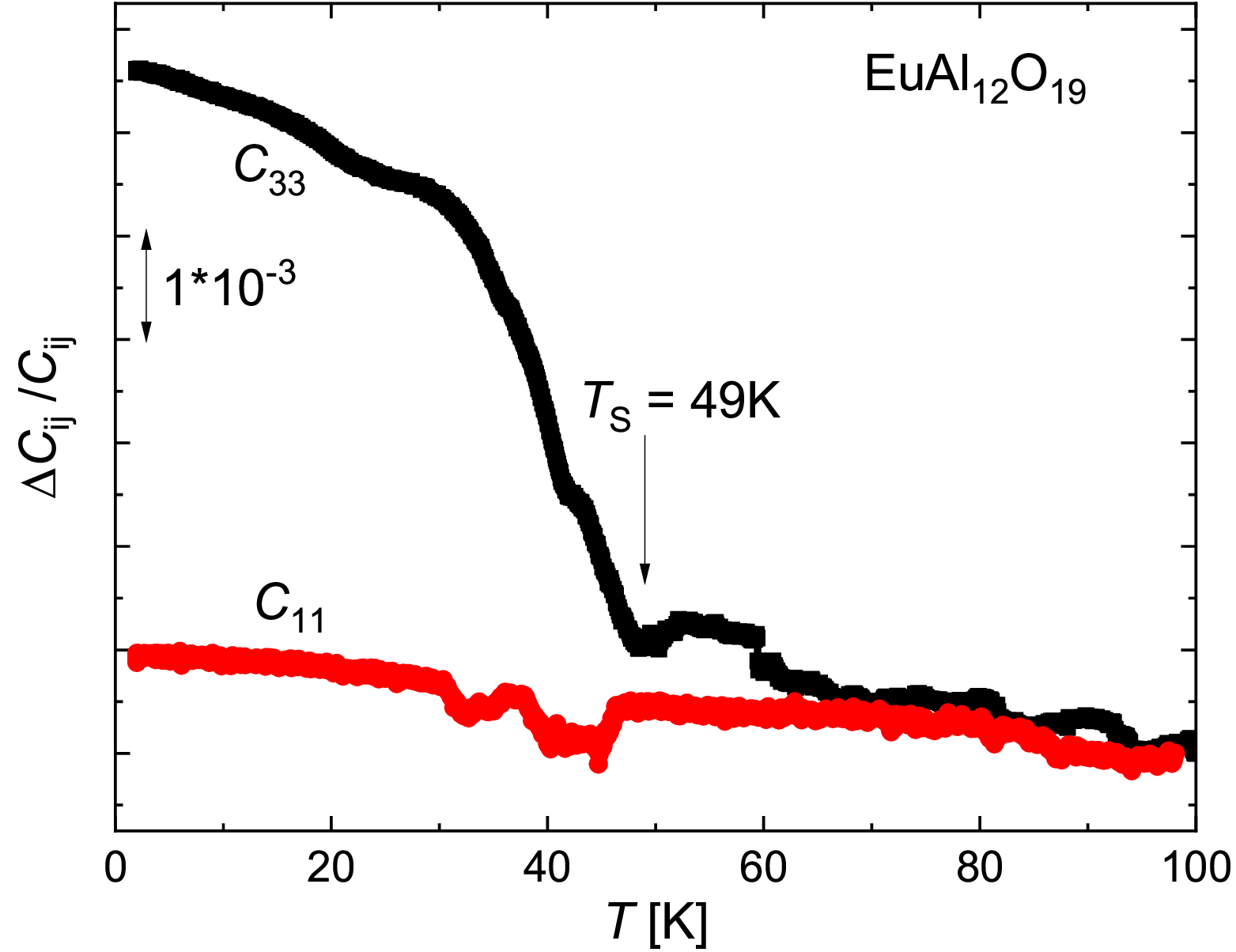}
		\caption{Relative changes of elastic constants C$_{11}$ and C$_{33}$ with temperature obtained from sound velocity measurements. The phase transition at $T_\mathrm{S}=49$\,K is indicated by a kink in the temperature dependence of the C$_{33}$ mode.}
		\label{accoustics} 
	\end{center}
\end{figure}

For further insights into the origin of the phase transition at $T_\mathrm{S}=49\,K$ in EuAl$_{12}$O$_{19}$, we measured the temperature dependence of sound velocity (Fig.~S\ref{accoustics}).
The single crystal of EuAl$_{12}$O$_{19}$ belongs to hexagonal symmetry and has 5 independent elastic constants (C$_{11}$, C$_{12}$, C$_{14}$ C$_{33}$, C$_{44}$). We focus here on the longitudinal modes C$_{11}$ and C$_{33}$. The temperature dependence of the longitudinal mode along the $c$ axis C$_{33}$ shows a kink at $T_\mathrm{S}=49$\,K and a hardening below this transition whereas the longitudinal mode along the $a$ axis C$_{11}$ shows a much weaker hardening. The hardening along the $c$ axis can be ascribed both to structural changes associated with the phase transition at $T_\mathrm{S}=49\,K$ and to the change of local structure associated with the progressive formation of antipolar correlations between neighboring electric dipoles.

\bibliography{EuAl12O19-ferroelectricity}